\documentclass[a4paper,11pt,oneside,fleqn]{article}

\usepackage[mathscr]{eucal}
\usepackage{amsmath,amssymb,amsthm}
\usepackage{graphicx}
\usepackage{caption}
\usepackage{subcaption}
\usepackage{color}

\usepackage{cite}
\usepackage{hyperref}
\hypersetup{colorlinks, linkcolor=blue, citecolor=blue, urlcolor=blue}

\allowdisplaybreaks

\makeatletter
\long\def\@makecaption#1#2{%
  \vskip\abovecaptionskip\footnotesize
  \sbox\@tempboxa{#1. #2}%
  \ifdim \wd\@tempboxa >\hsize
    #1. #2\par
  \else
    \global \@minipagefalse
    \hb@xt@\hsize{\hfil\box\@tempboxa\hfil}%
  \fi
  \vskip\belowcaptionskip}
\makeatother

\newcommand{\todo}[1][\null]{\ensuremath{\clubsuit}}

\newcommand{\noprint}[1]{}
\newcommand{\checked}[1][\null]{\ensuremath{\boldsymbol{\surd}}}

\newcommand{\const}{\mathop{\rm const}\nolimits}

\DeclareMathOperator\dry{dry}
\DeclareMathOperator\wet{wet}

\newtheorem*{problem*}{Problem}
{\theoremstyle{definition}

\newtheorem*{remark*}{Remark}
}

\begin{document}

\par\noindent {\LARGE\bf
A well-balanced meshless tsunami propagation\\ and inundation model
\par}

{\vspace{4mm}\par\noindent {\bf R\"udiger Brecht$^\dag\, ^\ddag$, Alexander Bihlo$^\dag$, Scott MacLachlan$^\dag$  and J\"orn Behrens$^\ddag$
} \par\vspace{2mm}\par}

{\vspace{2mm}\par\noindent {\it
$^{\dag}$~Department of Mathematics and Statistics, Memorial University of Newfoundland,\\ St.\ John's (NL) A1C 5S7, Canada
}}
{\vspace{2mm}\par\noindent {\it
$^{\ddag}$~Department of Mathematics, Universit\"at Hamburg, Bundesstra\ss e 55, Hamburg 20146, Germany and
CEN -- Center for Earth Systems Research and Sustainability, Universit\"at Hamburg,\\ Grindelberg 5, Hamburg 20144, Germany
}}

{\vspace{2mm}\par\noindent {\it
\textup{E-mail:} rbrecht@mun.ca, abihlo@mun.ca, smaclachlan@mun.ca, joern.behrens@uni-hamburg.de
}\par}

\vspace{4mm}\par\noindent\hspace*{10mm}\parbox{140mm}{\small
We present a novel meshless tsunami propagation and inundation model. We discretize the nonlinear shallow-water equations using a well-balanced scheme relying on radial basis function based finite differences. The inundation model relies on radial basis function generated extrapolation from the wet points closest to the wet--dry interface into the dry region. Numerical results against standard one- and two-dimensional benchmarks are presented.
}\par\vspace{4mm}

\section{Introduction}

The shallow-water equations are a simplified, albeit important, model
in geophysical fluid mechanics, playing a central role in both atmospheric and ocean sciences. In ocean dynamics, the shallow-water equations govern the evolution of long waves and, thus, are a particularly suitable model for tsunami propagation~\cite{goto97a,hari08a,leve11Ay,tito97Ay}.
An important characterization of a tsunami event is the involvement of vastly different scales, with wave propagation happening over thousands of kilometers in the open ocean and inundation taking place on coastlines with diverse features on the scale of just a few meters. This presence of vastly different scales has justified the extensive use of unstructured meshes and adaptive mesh refinement for the numerical solution of the shallow-water equations~\cite{hari08a,leve11Ay}.

Current numerical methodologies for the solution of the shallow-water
equations for tsunami modeling include finite-difference
methods~\cite{goto97a,tito98a,tito97Ay}, finite-volume methods~\cite{leve11Ay}, and finite-element and discontinuous Galerkin methods~\cite{hari08a,vate15a}. The application of so-called \textit{meshless methods} to the problem of tsunami modeling has received considerably less attention. 

One primary appeal of general meshless methods is that they do not
rely on a predefined topologically connected mesh but, rather, operate
on an (in principle) arbitrary collection of nodes where the numerical
solution of the model problem is sought. For this reason, meshless
methods are, by design, suitable for problems that benefit from
variable spatial resolution. Several meshless methods have been
proposed already for the shallow-water equations, such as those based
on smoothed particle hydrodynamics~\cite{dele10a,xia13a} or radial
basis functions~\cite{bihl17a,flye09Ay,flye12a,hon99a}. To the best of
our knowledge, the shallow-water equations with variable bottom
topography have not been considered extensively within the general
meshless methodology, as it is quite challenging to preserve the
inherent balance between the bottom topography source term and the flux term in the momentum equations. For a review on these difficulties and possible solutions, we refer to~\cite{bihl17a,xia13a}.

The purpose of this paper is to develop a meshless tsunami model based on radial basis functions generated finite differences (RBF-FD). The RBF-FD method was first introduced in~\cite{tols00a} and has since seen an extensive development both theoretically and with regards to different fields of applications, in particular in the geosciences~\cite{forn15a}. This method has methodology similar to classical finite differences, yet it can be used on both arbitrary nodal layouts and general smooth embedded manifolds~\cite{shan15a}, such as on the sphere~\cite{flye12a}. This makes RBF-FD a natural method for both far- and near-field simulations of tsunamis. Moreover, we demonstrate in this paper that RBF-based extrapolation is also a suitable method for the simulation of tsunami inundation.

The further organization of the paper is the following. In
Section~\ref{sec:ShallowWaterDiscretization}, we present the
shallow-water equations and discuss the well-balanced meshless RBF-FD
methodology employed for discretizing
them. Section~\ref{sec:InundationAlgorithm} is devoted to the
description of the RBF-based inundation algorithm. In
Section~\ref{sec:NumericalSimulationsTsunami}, we present the results
of various standard benchmark tests for the one- and two-dimensional
shallow-water equations aiming at assessing the quality of the newly
proposed method. The conclusions of the paper as well as a discussion
of necessary further developments are found in Section~\ref{sec:ConclusionsTsunami}.

\section{Well-balanced meshless discretization\texorpdfstring{\\}{ }of the shallow-water equations}\label{sec:ShallowWaterDiscretization}

In this section, we introduce the shallow-water equations and describe a well-balanced meshless method for their discretization.

\subsection{The shallow-water equations}

We consider the following generalized 2-dimensional conservation law
\begin{equation}\label{eq:ShallowWaterEqs}
 \boldsymbol{\rho}_t+\mathbf{F}_x+\mathbf{G}_y=\mathbf{S}.
\end{equation}
In order to define shallow-water equations, we specify (following~\cite{pedl87Ay}): $\rho=(h,hu,hv)^{\rm T}$ is the transport vector of total mass and horizontal momentum, $\mathbf{F}=(hu,hu^2+gh^2/2,huv)^{\rm T}$ and $\mathbf{G}=(hv,huv,hv^2+gh^2/2)^{\rm T}$ are the associated flux vectors, and $\mathbf{S}=(0,-ghb_x,-ghb_y)^{\rm T}$ is the source term. The height of a constant density water column is denoted by $h=h(t,x,y)$, the two-dimensional, vertically averaged fluid velocity is denoted by $(u,v)^{\rm T}=(u(t,x,y),v(t,x,y))^{\rm T}$, the sea bottom topography is $b=b(x,y)$, and $g$ is the gravitational constant. For the sake of brevity, we use subscripts to denote the partial derivatives with respect to $t$, $x$ and $y$, i.e.\ $h_t=\partial h/\partial t$, etc.

We point out that the form of the shallow-water equations used here
does not include any bottom friction, although such friction terms can
be introduced  with the aim of obtaining more accurate inundation
results~\cite{kais11a}. Here, we present mostly canonical benchmarks,
which are typically tested without the use of bottom friction and, so,
we also exclude this in our model. The inclusion of bottom friction as well as the application to more elaborate test cases will be the subject of future investigations.

\subsection{RBF-FD discretization}\label{sec:RBFFDDiscretization}

We discretize~\eqref{eq:ShallowWaterEqs} using the RBF-FD
method. Radial basis function based discretizations have seen a rapid
development over the past 20 years, both for the global formulation
(which is akin to the pseudospectral method~\cite{forn04a}) and for the local RBF-FD formulation, see e.g.~\cite{forn15a,forn15b} for recent reviews. The RBF methodology has also already been used for solving the shallow-water equations, both in planar geometry and on the surface of the sphere~\cite{flye09Ay,flye12a,hon99a,wong02a,zhou04a}. 

The discretization of the partial derivative operators
in~\eqref{eq:ShallowWaterEqs} within the framework of the RBF method
proceeds as follows. Consider a set of $N$ nodes, $\mathbf{x}_i\in\mathbb{R}^2$, covering the spatial domain~$\Omega\subseteq \mathbb{R}^2$ such that no two points coincide. Given the values of the field functions~$f\in\{h,u,v\}$ at those nodes, i.e.\ $f_j=f(\mathbf{x}_j)$, we approximate the action of a linear differential operator~$\mathcal L$ on $f$ at $\mathbf{x}_i$ by
\begin{equation}\label{eq:NodalDerivativeApproximation}
 \mathcal L f|_{\mathbf{x}=\mathbf{x}_i} \approx \sum_{j=1}^Nw^{\mathcal L}_{ij}f_j.
\end{equation}
In other words, the action of $\mathcal L$ on $f$ at $\mathbf{x}_i$ is approximated as a weighted linear sum of the values of~$f$ at the~$N$ nodes $\mathbf{x}_j$. Within the context of RBF methods, the weights~$w^{\mathcal L}_{ij}$ are found by enforcing~\eqref{eq:NodalDerivativeApproximation} to be exact when evaluated for radial basis functions~$\phi_k(\mathbf{x})=\phi(||\mathbf{x}-\mathbf{x}_k||)$ centered at $\mathbf{x}_k$, i.e.\
\begin{equation}\label{eq:RBFDerivativeSystem}
\mathcal L\phi_k(\mathbf{x}_i)=\sum_{j=1}^Nw^{\mathcal L}_{ij}\phi_k(\mathbf{x}_j),\quad k=1,\dots, N,
\end{equation}
which constitutes a linear system for $w_{ij}^{\mathcal L}$, $1\leq j
\leq N$,  at each node $\mathbf{x}_i$. In the RBF-FD method, one assumes that only points close to the node $\mathbf{x}_i$ contribute to the approximation of the derivative at $\mathbf{x}_i$ and, thus, most of the $N$ weights $w^{\mathcal L}_{ij}$ vanish. In practice, this is done by determining the $n$ nearest neighbors of $\mathbf{x}_i$, where typically $n\ll N$. Solving the resulting (small) linear system~\eqref{eq:RBFDerivativeSystem} at each node $\mathbf{x}_i$ allows one to compose the (sparse) differentiation matrix $W^{\mathcal L}=(w_{ij}^{\mathcal L})$, such that the action of the linear differential operator~$\mathcal L$ on the field functions $f$ at all nodal points $\mathbf{x}=(\mathbf{x}_1,\dots,\mathbf{x}_N)^{\rm T}$ can be approximated as $\mathcal L\mathbf{f}\approx W^{\mathcal L}\mathbf{f}$, where $\mathbf{f}=(f(\mathbf{x}_1),\dots,f(\mathbf{x}_N))^{\rm T}$. In the following, we have $\mathcal L\in\{\partial/\partial x,\partial/\partial y\}$. Note that it is also customary to include certain low-degree polynomials while solving~\eqref{eq:RBFDerivativeSystem}, as with a pure RBF basis it is not possible to obtain the correct derivatives of constants, linear functions, etc. For further details, see e.g.~\cite{bayo10a,forn13a,forn15a}.

Several RBFs are typically used, with the multiquadric, $\phi(r)=\sqrt{1+(\epsilon r)^2}$, and the Gaussian RBF, $\phi(r)=\exp(-(\epsilon r)^2)$ being amongst the most popular choices. The parameter~$\epsilon$ is called the \textit{shape parameter} as it governs the flatness of the RBF. In the following we will work with the Gaussian RBF, since for hyperbolic PDEs the RBF-FD method typically requires the use of hyperdiffusion for stabilization which is most easily accomplished using the Gaussian RBF~\cite{forn11a}. 

When implementing a numerical scheme for the shallow-water equations for ocean modeling, the presence of a nonflat sea bottom topography generally presents a challenge for both meshbased and meshless numerical schemes. More specifically, preserving the so-called \textit{lake at rest solution} is a nontrivial, but crucial endeavor, since the violation of the lake at rest solution typically leads to the stimulation of spurious numerical waves that can render the correct simulation of the actual physical waves extremely challenging. There has been a considerable body of literature devoted to the construction of so-called \textit{well-balanced} numerical schemes for the shallow-water equations, which are schemes that can preserve the lake at rest solution numerically, see~\cite{audu04a,gall03a,kurg07a,leve98a,vate15a,xia13a} for some examples for such well-balanced schemes. 

Most recently, in~\cite{bihl17a}, a unifying strategy was proposed for developing general well-balanced meshbased and meshless schemes which is in particular suitable for the RBF-FD methodology. For the sake of completeness of the present exposition, we briefly review the key idea of~\cite{bihl17a} here for the case of the one-dimensional form of the shallow-water equations,
\[
 h_t+(hu)_x=0,\qquad (hu)_t+\left(hu^2+\frac12gh^2\right)_x=-ghb_x.
\]
The two-dimensional case is treated analogously by enforcing the well-balanced condition given below for both the $x$- and $y$-derivatives.

As $u=0$ in the lake at rest solution, any well-balanced discretization has to preserve the identity
\begin{equation}\label{eq:ShallowWaterBalanceEquation}
 \frac12\partial_xh^2=-h\partial_xb,
\end{equation}
numerically in the case that $h+b=c$, for $c=\const$. It is found
in~\cite{bihl17a} that this is in general only possible if one
discretizes the balance
equation~\eqref{eq:ShallowWaterBalanceEquation}, at each point
$\mathbf{x}_i$, so that
\[
 \frac12\left(\mathrm{D}^{\rm f}_xh^2\right)_i=-\overline{h}_i\left(\mathrm{D}^{\rm s}_x(c-h)\right)_i,
\]
for any mesh function, $h$, and constant, $c$, where $\mathrm{D}_x^{\rm f}$ and $\mathrm{D}_x^{\rm s}$ are the discrete approximations to the spatial first derivative operators in the flux and source terms of the shallow-water equations, respectively, and $\overline{h}_i=\sum_{j=1}^nm_{ij}h_j$ is a specific, consistent average (meaning that $\sum_{j=1}^nm_{ij}=1$) of the total water height $h$ over the stencil of $\mathbf{x}_i$.

As noted in~\cite{bihl17a}, the above equality implies that
\begin{equation}\label{eq:ConditionsOnWellBalancedDerivatives}
 \left(\mathrm{D}^{\rm s}_xc\right)_i=0,\qquad \frac12\left(\mathrm{D}^{\rm f}_xh^2\right)_i=\overline{h}_i\left(\mathrm{D}^{\rm s}_xh\right)_i,
\end{equation}
must hold at all nodes $\mathbf{x}_i$. The first condition requires
the consistent (exact) derivative of constants by the derivative
operator $\mathrm{D}^{\rm s}_x$ (which in our case of RBF-FD based
derivatives requires the inclusion of zero-degree polynomials in the
basis), and the second condition can be satisfied provided we interpret $\mathrm{D}^{\rm f}_xh^2$ as a bilinear form, i.e.\
\[
\left(\frac12\mathrm{D}_x^{\rm f}h^2\right)_i=\frac12\mathbf{h}^{\rm T}W^{\rm f}_i\mathbf{h}.
\]
This procedure defines a third order differentiation tensor, $\mathbf{W}^{\rm f}$, with the matrix $W^{\rm f}_i$ being its $i$th slice. 

For the approximation of the source derivative, $\mathrm{D}_x^{\rm s}$, we write
\[
 (\mathrm{D}^{\rm s}_xh)_i=(\mathbf{w}^{\rm s}_i)^{\rm T}\mathbf{h},
\]
where $\mathbf{w}^{\rm s}_i=(W_{ij}^{\rm s})_{1\leqslant j\leqslant n}$ is the $i$th row of the associated differentiation matrix~$W^{\rm s}$. In a similar manner, we can write $\mathbf{m}_i = (m_{ij})_{1\leqslant j\leqslant n}$ for the $i$th row of the averaging matrix~$\mathrm M$. The second condition of~\eqref{eq:ConditionsOnWellBalancedDerivatives} then naturally translates to
\[
\frac12\mathbf{h}^{\rm T}W^{\rm f}_i\mathbf{h}=(\mathbf{m}^{\rm T}_i\mathbf{h})((\mathbf{w}^{\rm s}_i)^{\rm T}\mathbf{h}),
\]
for all $i$, which (taking $W^{\rm f}_i$ to be symmetric) requires that
\begin{equation}\label{eq:ConditionsOnWellBalancedDerivativesMatrix}
W^{\rm f}_i=\mathbf{m}_i\left(\mathbf{w}^{\rm s}_i\right)^{\rm T} + \mathbf{w}^{\rm s}_i\mathbf{m}_i^{\rm T},
\end{equation}
holds at all nodes~$\mathbf{x}_i$. 

Practically speaking, one is thus free to choose the averaging matrix~$\mathrm M$ and the derivative matrix $\mathrm{W}_x^{\rm s}$ corresponding to the derivative approximation used in the source term and then Eq.~\eqref{eq:ConditionsOnWellBalancedDerivativesMatrix} prescribes how to choose the weights in the flux derivative~$\mathrm{D}_x^{\rm f}h^2$, given through the derivative tensor $\mathbf{W}^{\rm f}$, so as to obtain a well-balanced scheme for the shallow-water equations. For a more in-depth discussion and results regarding the consistency of the resulting discrete derivative operators, consult~\cite{bihl17a}.

We use the RBF-FD method for discretizing the two-dimensional shallow-water equations~\eqref{eq:ShallowWaterEqs}, invoking condition~\eqref{eq:ConditionsOnWellBalancedDerivativesMatrix} to guarantee that the resulting meshless scheme will be well-balanced. For the time-stepping, the second-order explicit midpoint scheme is used. It is pointed out in~\cite{forn11a} that the application of the RBF-FD method to purely convective PDEs is prone to numerical instability as the eigenvalues of the associated derivative matrices tend to scatter to the complex right-half plane. As a remedy, the inclusion of hyperdiffusion was proposed, which we have included in our discretization as well. In other words, instead of solving the shallow-water equations~\eqref{eq:ShallowWaterEqs}, we solve
$
 \boldsymbol{\rho}_t+\mathbf{F}_x+\mathbf{G}_y=\mathbf{\tilde S},
$
where $\mathbf{\tilde S}$ is a modified source term of the form $\mathbf{\tilde S}=\mathbf{S}+\mathbf{D}$, where
\[
\mathbf{D}=(-1)^{\ell+1}\nu\Delta^\ell\left(0,hu,hv\right)^{\rm T},\qquad \ell\geqslant1
\]
with $\nu$ being the diffusion parameter and $\Delta=\partial^2/\partial x^2+\partial^2/\partial y^2$ being the two-dimensional Laplacian operator. If the underlying RBF is the Gaussian RBF, then $\Delta^\ell\phi(r)=\epsilon^{2\ell}p_\ell(r)\phi(r)$, with $p_\ell(r)$ being computable through the recurrence relation
\begin{equation}\label{eq:RecurrenceRelation}
 p_0(r)=1,\quad p_1(r)=4(\epsilon r)^2-4,\quad p_{\ell+1}(r)=4((\epsilon r)^2-2\ell-1)p_\ell(r)-16\ell^2p_{\ell-1}(r),\quad \ell\geqslant1.
\end{equation}
The polynomials~$p_\ell(r)$ are related to the Laguerre orthogonal
polynomials, see~\cite{forn11a} for further details. Below, we use
$\ell=2$, unless otherwise noted.

To summarize, we present below the algorithm used at every time step.
The initialization phase of the algorithm uses the RBF-FD methodology
to define the averaging matrix, $\mathrm M$, source derivative, $\mathrm W_i^{\rm s}$,
as well as the discrete approximation of the hyperdiffusion operator,
$\mathbf{D}$.  Additionally, we precompute $b^{\rm d}_x$ and $b^{\rm
  d}_y$, the RBF-FD derivatives of the bottom topography function,
$b(x,y)$.  In the algorithm below, we use $\mathrm D_x^{\rm d}$ and $\mathrm D_y^{\rm
  d}$ to denote the discrete derivative
operators of the well-balanced scheme, with the discrete
hyperdiffusion operator, $\mathrm D_{\Delta^2}^{\rm d}$, computed using
\eqref{eq:RecurrenceRelation}.  At each node, $i$, we store unknowns
$h_i$, $(hu)_i$, and $(hv)_i$; the products and ratios of vectors $h$,
$hu$, and $hv$ given in the algorithm below are computed componentwise.  We note
that the use of the inundation algorithm presented below ensures that
no division by zero is ever performed.  In what follows, at each time
step, we make use of the natural partitioning of the set of all nodes,
$X = \left\{\mathbf{x}_i\right\}_{i=1}^N$, into sets of wet and dry
  points,
\[
 X_{\rm wet}=\{\mathbf{x}_i\in X : h_i> \delta\},\quad X_{\rm dry}=X\backslash X_{\rm wet},
\]
where $\delta\in\mathbb{R}$ is a user-defined (small) positive
parameter.  We note that this partitioning is updated after every
stage of the time-step, but we supress superscripts or subscripts
denoting the time-step and stage to simplify notation.

\begin{description}
	\item[For every timestep:] ~
	\begin{description}

\item[Compute first stage of explicit midpoint rule:]~\\
Compute $hu^2=h\cdot(\frac{hu}{h})^2, hv^2=h\cdot(\frac{hv}{h})^2,
hvu=h\cdot\frac{hv}{h}\cdot\frac{hu}{h}
$
and
 $\bar h = \mathrm Mh$, then set $(\mathfrak{h}_x)_i=(\bar h)_i\cdot (\mathrm
D_x^{\rm d} h)_i$ and $(\mathfrak{h}_y)_i=(\bar h)_i\cdot (\mathrm
D_y^{\rm d} h)_i$ for all nodes $\mathbf{x}_i \in X_{\rm wet}$.
\begin{align*}
\widetilde h &= h - \frac{\Delta t} 2 \big( \mathrm D_x^{\rm d}(hu)+\mathrm D_y^{\rm d}(hv)\big)
\\
\widetilde{hu}&= hu-\frac{\Delta t} 2 \big(\mathrm D_x^{\rm d}(hu^2)+ g \mathfrak{h}_x + \mathrm D_y^{\rm d}(hvu) + g\bar h b^{\rm d}_x -\eta\mathrm D_{\Delta^2}^{\rm d}(hu) \big) 
\\
\widetilde{hv}&= hv-\frac{\Delta t} 2 \big( \mathrm D_y^{\rm d}(hv^2)+ g \mathfrak{h}_y + \mathrm D_x^{\rm d}(hvu) + g\bar h b^{\rm d}_y -\eta \mathrm D_{\Delta^2}^{\rm d}(hv) \big) 
\end{align*}
\item[Apply boundary conditions.]~
\item[Apply inundation algorithm.]~
\item[Compute second stage of explicit midpoint method:] ~\\
Compute $\widetilde{hu^2}=\widetilde{h}\cdot(\frac{\widetilde{hu}}{\widetilde{h}})^2, \widetilde{hv^2}=\widetilde{h}\cdot(\frac{\widetilde{hv}}{\widetilde{h}})^2,
\widetilde{hvu}=\widetilde{h}\cdot\frac{\widetilde{hv}}{\widetilde{h}}\cdot\frac{\widetilde{hu}}{\widetilde{h}}
$
and
$\overline{\tilde{h}} = \mathrm M\tilde{h}$, then set $(\widetilde{\mathfrak{h}_x})_i=(\overline{\tilde{h}})_i\cdot (\mathrm D_x^{\rm d} \widetilde{h})_i$ and $(\widetilde{\mathfrak{h}_y})_i=(\overline{\tilde{h}})_i\cdot (\mathrm D_y^{\rm d} \widetilde{h})_i$ for all nodes $\mathbf{x}_i \in X_{\rm wet}$.
\begin{align*}
h&\leftarrow h -\Delta t \big(\mathrm D_x^{\rm d}(\widetilde{hu})+ \mathrm D_y^{\rm d}(\widetilde{hv})\big)
\\
hu&\leftarrow hu-\Delta t \Big(
\mathrm D_x^{\rm d}(\widetilde{hu^2})+ g \widetilde{\mathfrak{h}_x} + \mathrm D_y^{\rm d}(\widetilde{hvu}) + g\overline{\tilde{h}} b^{\rm d}_x -\eta \mathrm D_{\Delta^2}^{\rm d}(\widetilde{hu})  \Big) 
\\
hv&\leftarrow hv-\Delta t \Big(
\mathrm D_y^{\rm d}(\widetilde{hv^2})+ g \widetilde{\mathfrak{h}_{y}} +\mathrm D_x^{\rm d}(\widetilde{hvu}) + g \overline{\tilde{h}} b^{\rm d}_y -\eta \mathrm D_{\Delta^2}^{\rm d}(\widetilde{hv})  \Big) 
\end{align*}
\item[Apply boundary conditions.]~
\item[Apply inundation algorithm.]~
	\end{description}
\end{description}

We note that, since the scheme is explicit, this allows a very
efficient implementation of the time-stepping using only sparse matrix
multiplications with $\mathrm M$ and $\mathrm W_{i}^{\rm s}$, along with componentwise
vector operations.  Furthermore, the fine-scale parallelism of these
operations allows a natural avenue for parallelization, on both
classical compute clusters and modern manycore and accelerated
architectures.

\section{Meshless inundation algorithm}\label{sec:InundationAlgorithm}

Of particular importance in a tsunami model is the treatment of the
wet--dry interface when the incoming wave hits the coastline. Several
algorithms have been proposed to deal with this moving boundary
condition in numerical models that solve the governing equations in
the strong form (for associated results for the governing equations in
the weak form, consult,
e.g.~\cite{bate99a,buny09a,vate15a}). In~\cite{tito95a}, the authors
use a finite-difference model with variable grid spacing near the
boundary that ensures existence of a shoreline boundary point on the surface of the beach at all times. Similarly, in~\cite{lyne02a}, based on the earlier work of~\cite{siel70a} for the shallow-water equations, a finite-difference model for the nonlinear Boussinesq equations with fixed grid spacing was used that handles the moving boundary by employing (one-dimensional) linear extrapolation of the wave run-up from the last wet points to the first dry points on the beach. This idea was re-considered in~\cite{fuhr08a} where true two-dimensional bilinear extrapolation was used to compute the run-up in a two-dimensional finite-difference model for the Boussinesq equations. 

Since our numerical scheme is based on the RBF methodology using the
strong form of the shallow-water equations, it is natural to use RBFs
also for the inundation model. This is, in particular, justified as
(multiquadric) RBFs were originally proposed by Hardy for
two-dimensional scattered data interpolation~\cite{hard71a} and later
found by Franke to be the most accurate of all the 29 methods tested in~\cite{fran82a}. Of further interest are the studies carried out in~\cite{forn07a} where the authors investigated the Runge phenomenon in the context of RBF interpolation. It has been found that the Runge phenomenon can be controlled by a suitable choice of the shape parameter in the RBF (with spatially varying shape parameters most favorable) and non-equally chosen data points, typically much better than the Runge phenomenon can be controlled in standard polynomial interpolation.

In light of the favorable performance of RBF-based interpolation
schemes, we propose to use the extrapolation technique
of~\cite{fuhr08a,lyne02a,siel70a} but instead of using
polynomial-based extrapolation, we use \textit{RBF-based
  extrapolation}.  We again make use of the partitioning of the set of
all nodes, $X$, into wet and dry points, as $X = X_{\rm wet} \cup
X_{\rm dry}$, with $X_{\rm wet}=\{\mathbf{x}_i\in X : h_i> \delta\}$. Since the shallow-water equations are not defined at the dry points (where $h_i$ is too small or negative), we need to extrapolate the wet values to the dry points to have a numerical solution defined at all points~$X$. The advantage of this procedure is that then the same derivative approximation can be invoked at all points, even those close to the wet--dry interface, where some of the nearest neighbors on which the RBF derivatives are defined will be dry points.

For each dry point immediately neighboring a wet point we find its $n_{\rm e}$ nearest wet points through which we define an RBF interpolant. Once the RBF interpolant based on the wet points is defined, we use it to extrapolate the field functions $u$, $v$ and $h+b$ to the dry point under consideration.  We found experimentally that the multiquadric RBF basis augmented with first degree polynomials (i.e.\ constants and the monomimals $x$ and $y$) is most suitable for the extrapolation procedure. Note that in order to preserve the lake at rest solution in the presence of dry points it is necessary to extrapolate $h+b$, not $h$ alone, to the dry points.
The extrapolation is done by the following algorithm:
\begin{description}
	\item[for each $\mathbf{x}^{\dry}\in X_{\dry}$ immediately neighboring a wet point]~
	\begin{description}
		\item[find $\{\mathbf{x}_{j_1}^{\wet}, \ldots, \mathbf{x}_{j_{nn}}^{\wet}\}$ nearest neighbours of $\mathbf{x}_k^{\dry}$ in $X_{\wet}$]
		\item[extrapolate $h(\mathbf{x}^{\dry}), hu(\mathbf{x}^{\dry}), hv(\mathbf{x}^{\dry})$:]
		\item for $f\in \{h, hu, hv\}$ solve the linear system for $w$
		\[
		\begin{pmatrix}
		\phi(r_{1,1})&…&\phi(r_{1,nn})&1&(\mathbf{x}_{j_1}^{\wet})^T\\
		\vdots &\ddots & &\vdots&\vdots\\
		\phi(r_{nn,1})&…&\phi(r_{nn,nn})&1&(\mathbf{x}_{j_{nn}}^{\wet})^T\\
		1&\ldots&1&0&0\\
		\mathbf{x}_{j_1}^{\wet}&\ldots&\mathbf{x}_{j_{nn}}^{\wet}&0&0
		\end{pmatrix}		
		\begin{pmatrix}
		w_1\\
		\vdots \\
		w_{nn}\\
		w_{nn+1}\\
		\mathbf{w}_{nn+2}
		\end{pmatrix}
		=
		\begin{pmatrix}
		f(\mathbf{x}^{\wet}_{j_1})\\
		\vdots\\
		f(\mathbf{x}^{\wet}_{j_{nn}})\\
		0\\
		0
		\end{pmatrix}
		\qquad \text{with } r_{i, l}=\|\mathbf{x}^{\wet}_{j_i}-\mathbf{x}^{\wet}_{j_l}\|
		\]		
		Set $\displaystyle f(\mathbf{x}^{\dry})=\sum_{i=1}^{nn} w_i \phi(r_i) + w_{nn+1}+\mathbf{w}_{nn+2}\cdot \mathbf{x}^{\dry}$
	\end{description}
\end{description}

Here, we note that the coefficient $\mathbf{w}_{nn+2}$ has the same dimension as $\mathbf{x}^{dry}$.

\section{Numerical simulations}\label{sec:NumericalSimulationsTsunami}

Having described the meshless discretization of the shallow-water equations~\eqref{eq:ShallowWaterEqs} and the associated inundation model, we now proceed to present the results of several classical benchmark tests for both the one-dimensional and two-dimensional form of the shallow-water equations.

\subsection{One-dimensional benchmarks}

We repeat here some of the one-dimensional tests carried out
in~\cite{vate15a}. In all experiments we compute the RBF-FD
approximation~$\mathrm{D}_x^{\rm s}$, obtained from the procedure
outlined in Section~\ref{sec:RBFFDDiscretization}. The RBF used is the
Gaussian RBF with shape parameter $\epsilon=0.1/\Delta x$, where
$\Delta x$ is the (uniform) nodal spacing. Eq.~\eqref{eq:RBFDerivativeSystem} is
solved based on the three nearest neighbors of each nodal point (which
includes each node itself). The averaging matrix~$\mathrm M$ needed
for~\eqref{eq:ConditionsOnWellBalancedDerivativesMatrix} is obtained
from a normalized Gaussian filter over the three nearest neighbors of
each node. 
With this, each time step of the explicit midpoint scheme can be implemented with
$\mathcal{O}(1)$ work per spatial mesh point.

\subsubsection{Lake at rest solution}

It was shown in~\cite{bihl17a} that the numerical scheme based on a discretization that respects~\eqref{eq:ConditionsOnWellBalancedDerivatives}--\eqref{eq:ConditionsOnWellBalancedDerivativesMatrix} is well-balanced. Since there the shallow-water equations were considered without an inundation model, we carry out a test case where part of the domain is initially dry, so as to check that the discretization remains well-balanced in the presence of wet--dry interfaces.

\begin{figure}[!ht]
\centering
\begin{subfigure}[b]{0.45\textwidth}
  \centering
  \includegraphics[width=\linewidth]{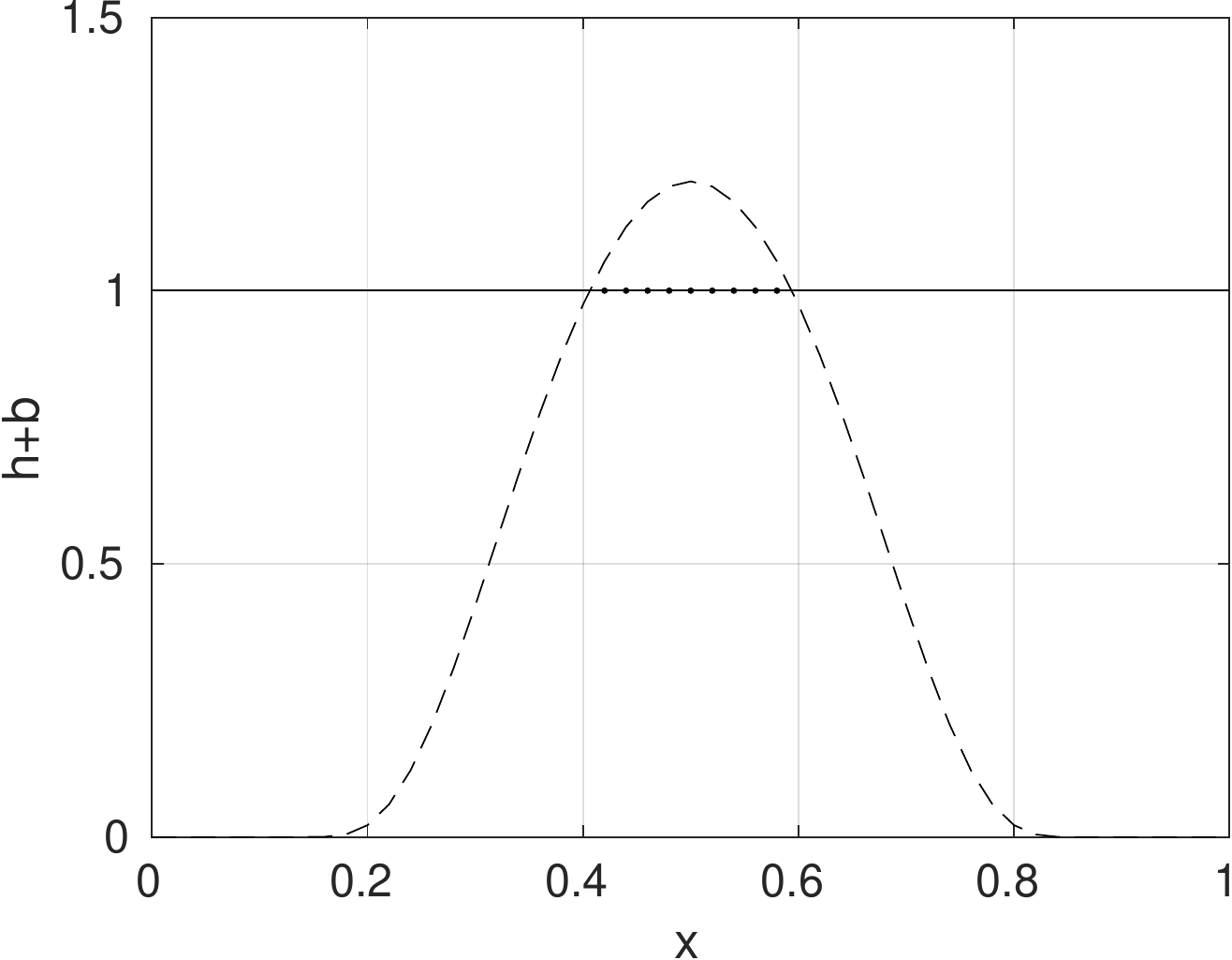}
\end{subfigure}\qquad
\begin{subfigure}[b]{0.45\textwidth}
  \centering
  \includegraphics[width=\linewidth]{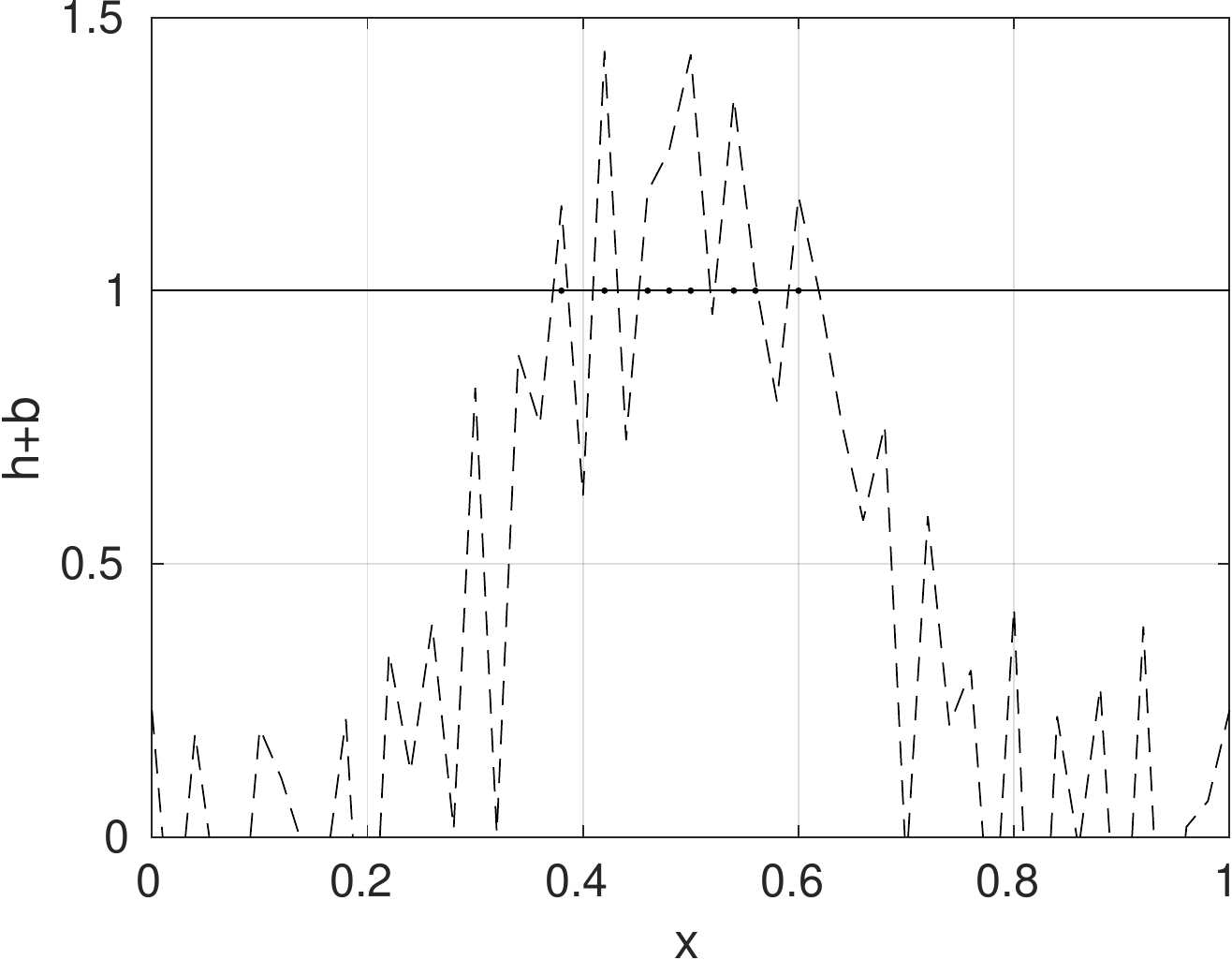}
\end{subfigure}
\caption{\footnotesize{Final surface elevation for the lake at rest solution with smooth (left) and noisy (right) bottom topography; bathymetry (dashed), surface elevation (solid) and interpolated nodes (solid with points).}}
\label{fig:LakeAtRest}
\end{figure}

\begin{figure}[!ht]
\centering
\begin{subfigure}[b]{0.45\textwidth}
  \centering
  \includegraphics[width=\linewidth]{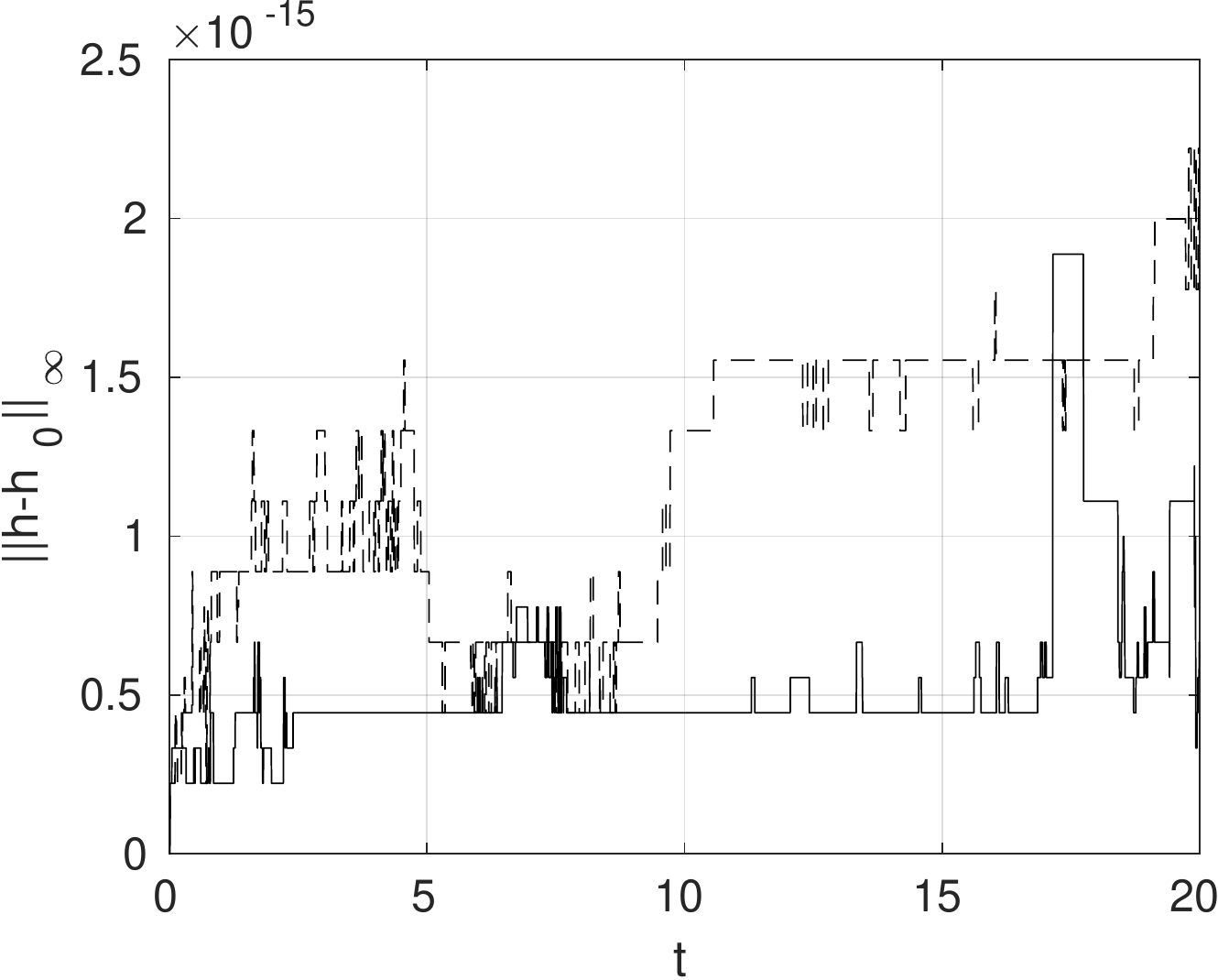}
\end{subfigure}\qquad
\begin{subfigure}[b]{0.45\textwidth}
  \centering
  \includegraphics[width=\linewidth]{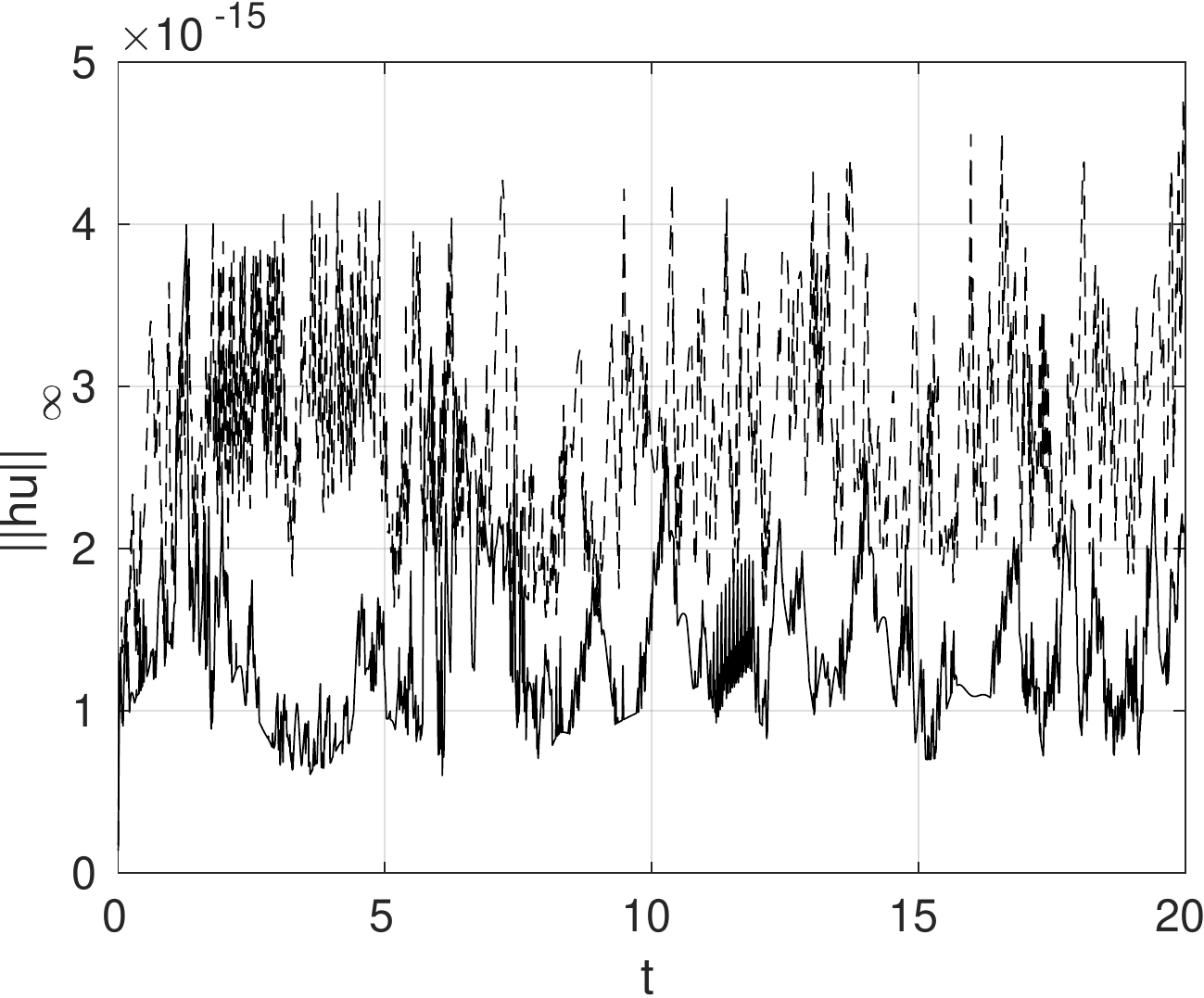}
\end{subfigure}
\caption{\footnotesize{$l_\infty$-error time series for the surface elevation (left) and momentum (right) for the lake at rest solution; smooth bathymetry (solid), noisy bathymetry (dashed).}}
\label{fig:LakeAtRestErrors}
\end{figure}

Specifically, we consider the domain $\Omega=[0,1]$, with the smooth bottom topography
\begin{equation}\label{bLakeatrestSmooth}
 b_{\rm s}=\left\{\begin{array}{cc} a\dfrac{\exp(-0.5/(r_m^2-r^2))}{\exp(-0.5/r_m^2)} & \textup{if}\quad r<r_m\\ 0 & \textup{otherwise}\end{array}\right.,
\end{equation}
where we set $r=|x-0.5|$, $a=1.2$ and $r_m=0.4$. We use $h=\max(0,1-b(x))$ and $u=0$ as initial conditions. A total of $N=50$ (regularly spaced) grid points were used, the time step is $\Delta t=0.002$ and the final integration time is $t=20$. The extrapolation tolerance is $\delta=0.0025$ with the three nearest neighbors being used for the extrapolation. The shape parameter for the RBF extrapolation is $\epsilon_{\rm e}=1$. Since realistic bathymetry is usually not smooth, in a second experiment, we add some noise on top of the smooth bottom topography $b_s$. In particular, we consider a noisy bathymetry of the form
\[
 b_{\rm n}=b_{\rm s}+\sum_{j=1}^3a_j\sin(16j\pi x+p_j),
\]
where $a=(0.1,0.2,0.3)$ and $p=(1.6,3.2,0.5)$. Reflecting boundary conditions were used for both experiments. The results of the two numerical experiments are depicted in Figures~\ref{fig:LakeAtRest} and~\ref{fig:LakeAtRestErrors}, which demonstrate that the extrapolating boundary conditions preserve the well-balanced properties of the meshless RBF-FD discretization. 

\subsubsection{Oscillatory flow in a parabolic basin}

We consider the oscillatory flow in a parabolic basin, which is
described by the following exact solution first reported
in~\cite{thac81a}. The domain for this problem is
$\Omega=[-5000,5000]$ with parabolic bottom topography $b=h_0(x/a)^2$,
where $a=3000$ and $h_0=10$. The initial conditions are chosen so that
the exact solution to the shallow-water equations for this benchmark is
\begin{align*}
 &h_{\rm a}(t,x)=h_0-\frac{B^2}{4g}(1+\cos2\omega t)-\frac{Bx}{2a}\sqrt{\frac{8h_0}{g}}\cos(\omega t),\quad
 u_{\rm a}(t,x)=\frac{Ba\omega}{\sqrt{2h_0g}}\sin\omega t
\end{align*}
where $\omega=\sqrt{2gh_0}/a$ and $B=5$. 

\begin{figure}[!ht]
\centering
\begin{subfigure}[b]{0.45\textwidth}
  \centering
  \includegraphics[width=\linewidth]{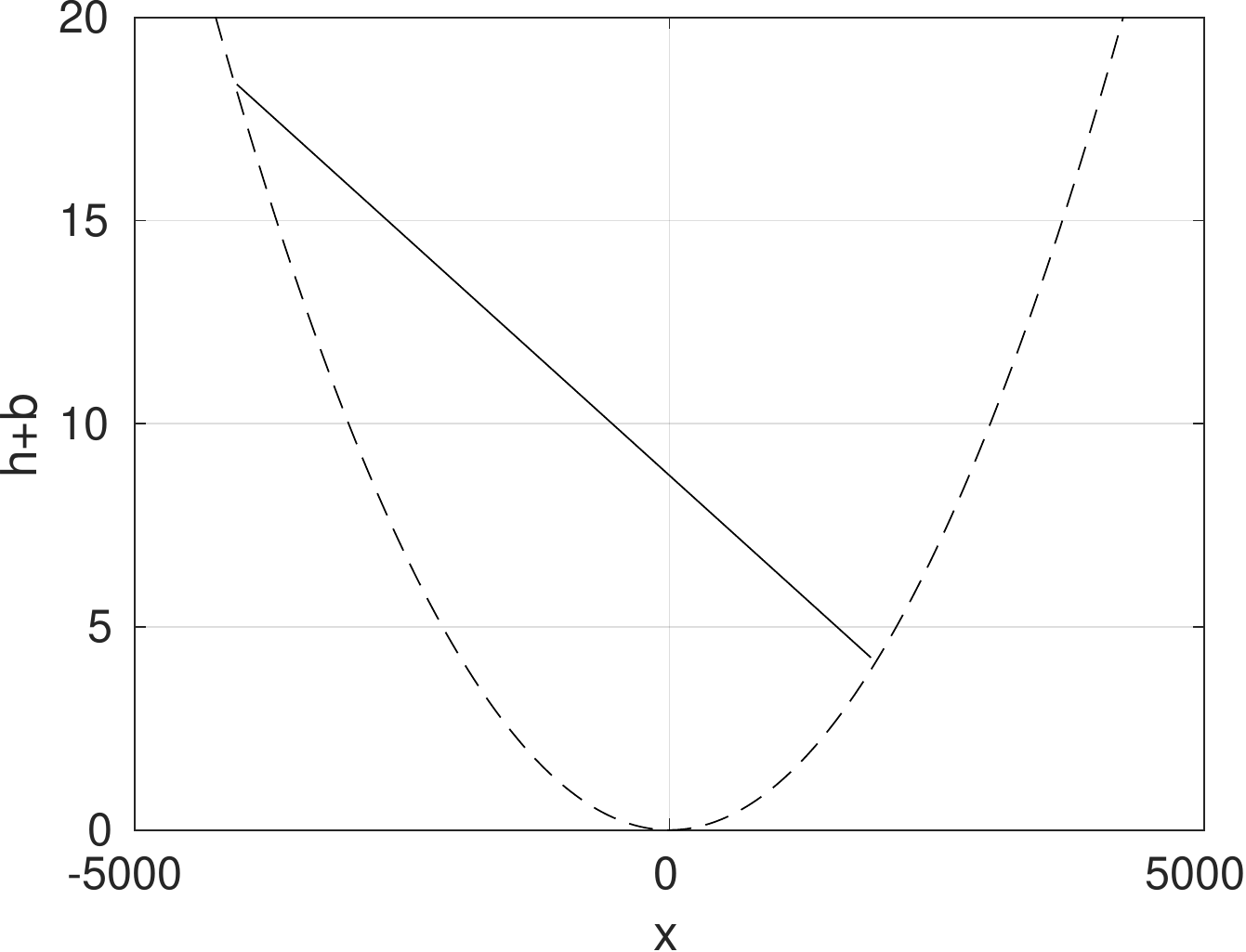}
\end{subfigure}\qquad
\begin{subfigure}[b]{0.45\textwidth}
  \centering
  \includegraphics[width=\linewidth]{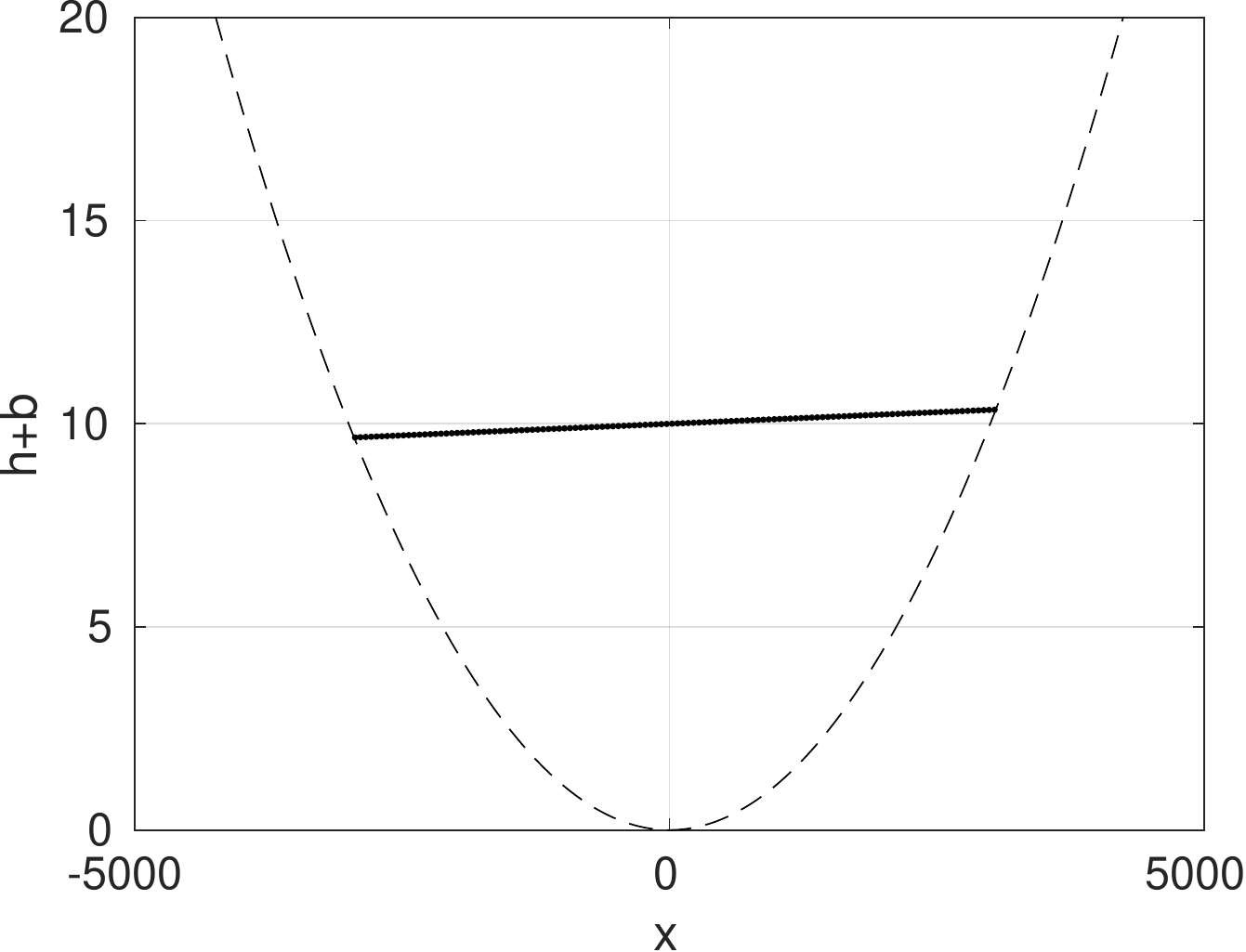}
\end{subfigure}\\
\begin{subfigure}[b]{0.45\textwidth}
  \centering
  \includegraphics[width=\linewidth]{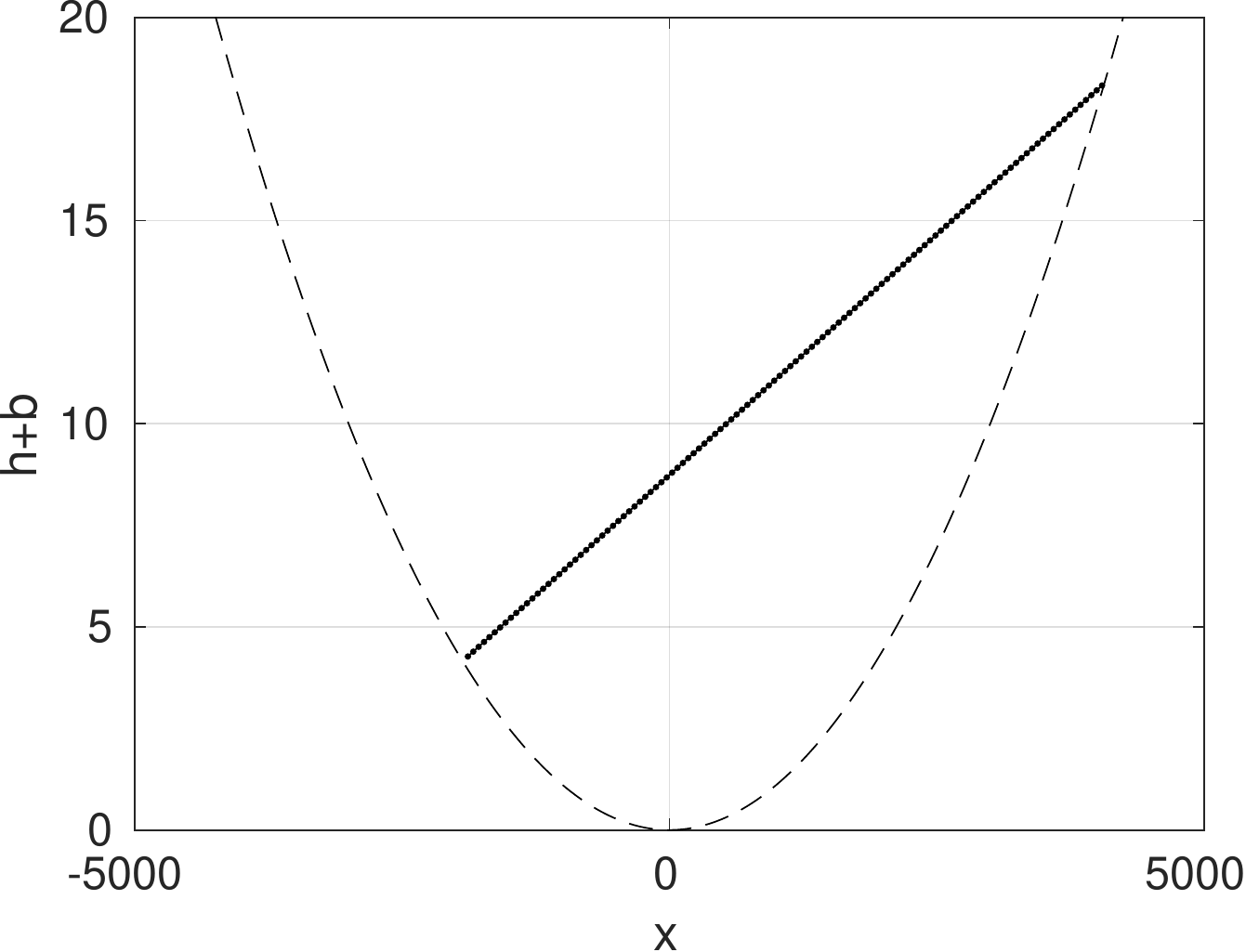}
\end{subfigure}\qquad
\begin{subfigure}[b]{0.45\textwidth}
  \centering
  \includegraphics[width=\linewidth]{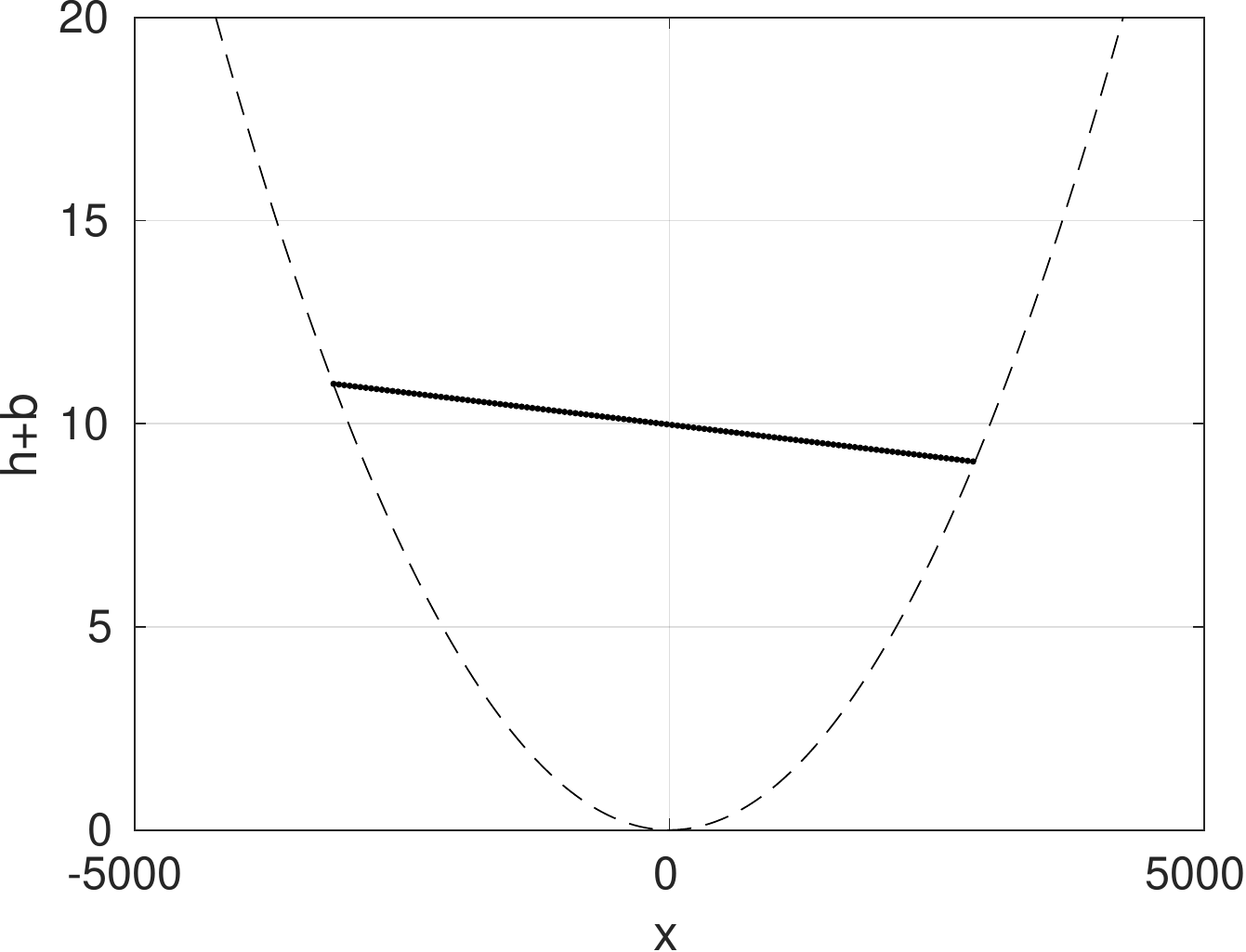}
\end{subfigure}
\caption{\footnotesize{Snapshots of the numerical solution (solid line) and analytical solution (dotted line) of the surface elevation for the parabolic bowl bathymetry (dashed line) at times $t=0$, $t=1000$, $t=2000$ and $t=3000$.}}
\label{fig:ParabolicBowlSolution}
\end{figure}

\begin{figure}[!ht]
\centering
\begin{subfigure}[b]{0.45\textwidth}
  \centering
  \includegraphics[width=\linewidth]{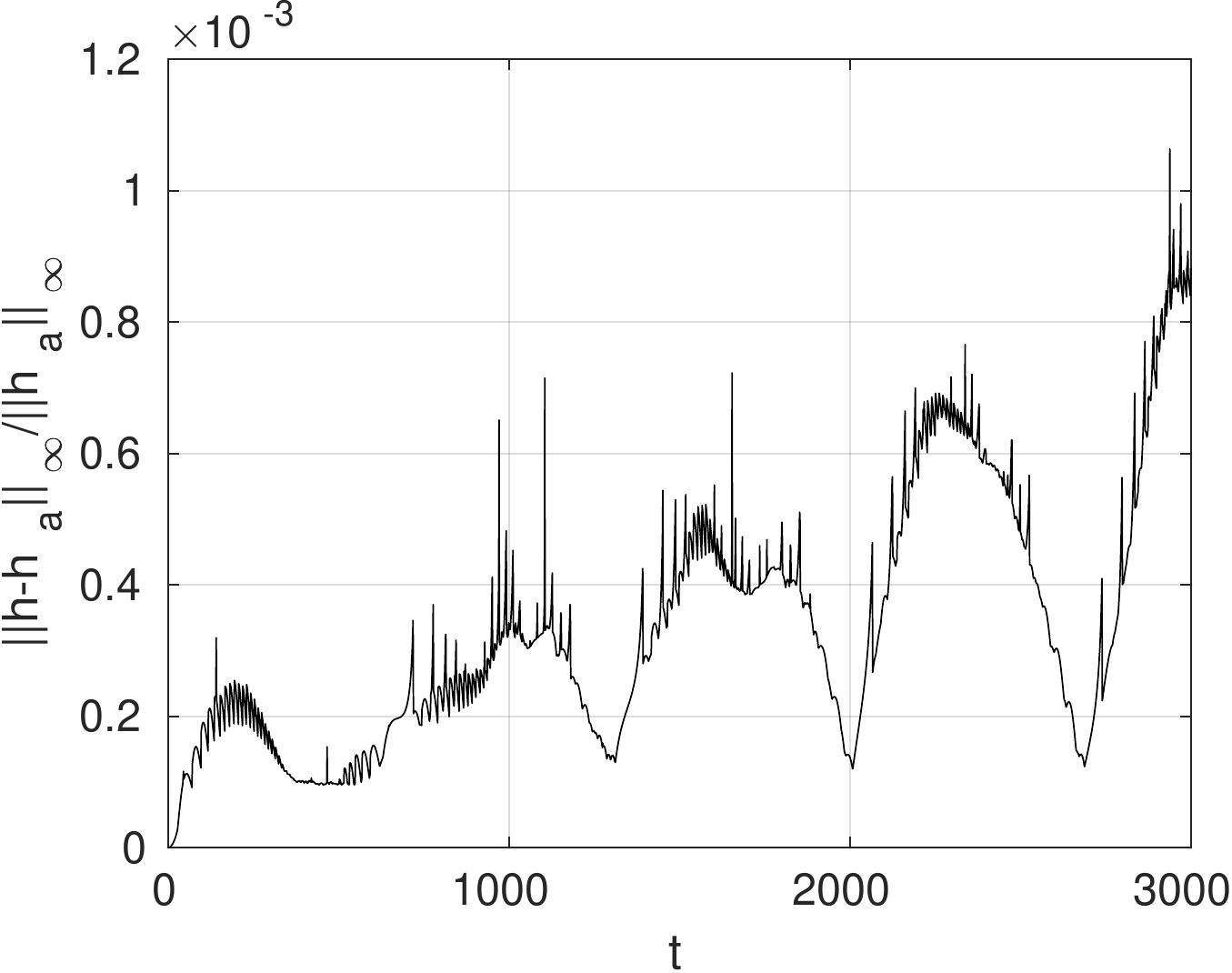}
\end{subfigure}\qquad
\begin{subfigure}[b]{0.45\textwidth}
  \centering
  \includegraphics[width=\linewidth]{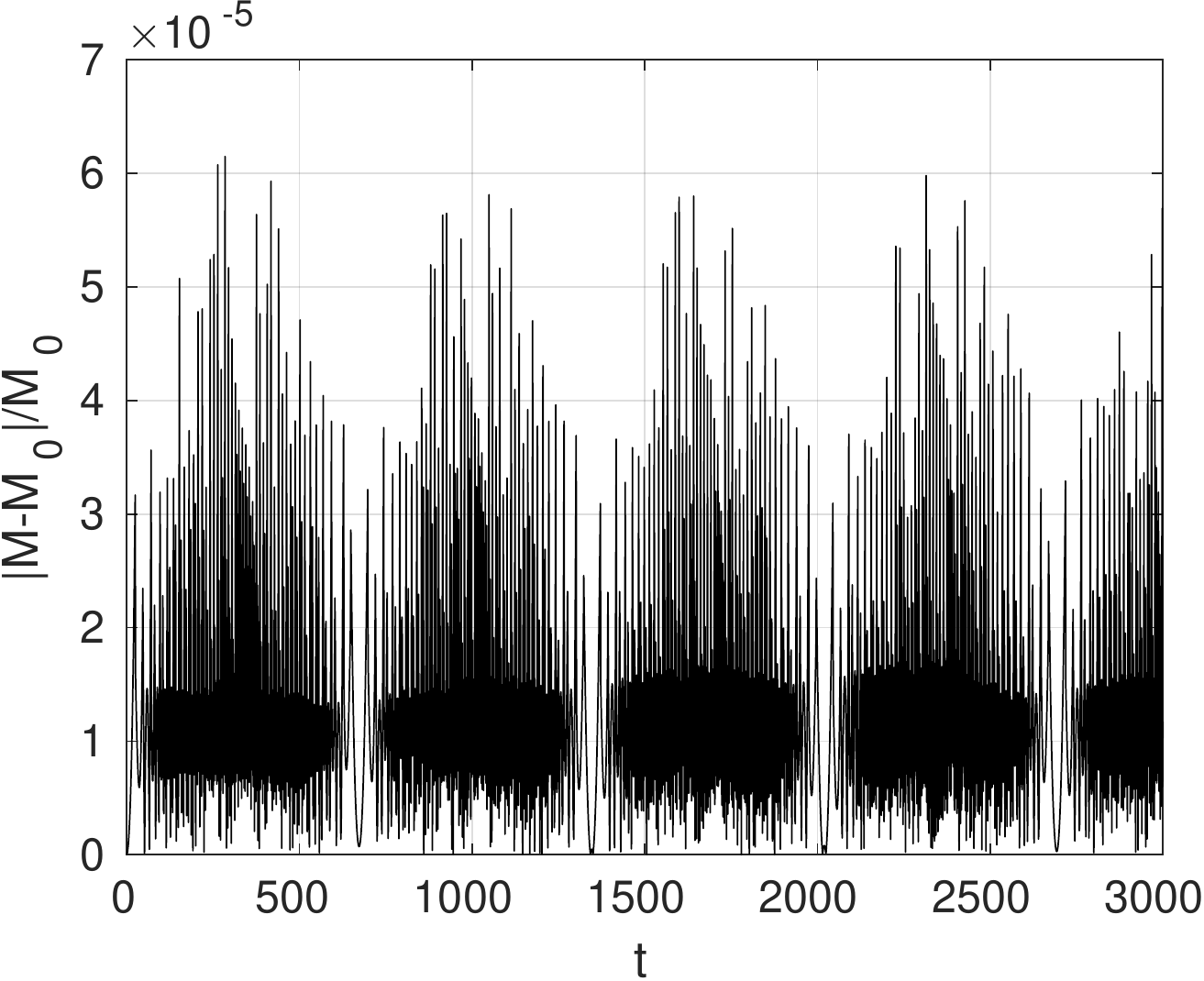}
\end{subfigure}
\caption{\footnotesize{$l_\infty$-error time series for the relative surface elevation error (left) and relative mass error (right).}}
\label{fig:ParabolicBowlErrors}
\end{figure}

In the first experiment we use $N=200$ equally spaced nodes and integrate the one-dimensional shallow-water equations up to $t=3000$ using the time step $\Delta t=1$. The extrapolation parameter is set to $\delta=0.01$. Since the water surface never reaches the boundaries of the domain, no specific boundary conditions have to be imposed. Snapshots of the numerical solution and the exact solution at times $t=0$, $t=1000$, $t=2000$ and $t=3000$ are depicted in Figure~\ref{fig:ParabolicBowlSolution}. The associated time series of the $l_\infty$ relative height and mass errors are shown in Figure~\ref{fig:ParabolicBowlErrors}.

Note that although our numerical scheme does not preserve mass by construction, the relative change in the total mass over time is small, bounded and oscillatory only. That is, no spurious trend in the mass is introduced by the numerical scheme and the inundation model.

To numerically verify the convergence of the numerical solution, we
next carry out a sequence of numerical integrations using
$N\in\{200,400,800,1600\}$ equally spaced points. The time steps are
halved each time the number of nodes is doubled. The final integration
time is again $t=3000$. In
Table~\ref{tab:ConvergenceStudyParabolicBowl}, we record the maximum
$l_\infty$ errors (relative height error, relative mass error,
absolute momentum error), occurred over the integration period. The
table shows also the experimental convergence rate
$\frac{\|e_c\|}{\|e_f\|}$, where $e_c$ is the error of the coarse and
$e_f$ of the fine grid.  Similar data is shown in
Table~\ref{tab:ConvergenceStudyParabolicBowl2}, for the $l_2$ norm; we
note that both measures of the error show similar overall convergence,
but the $l_2$ norm shows steadier convergence rates than the maximum
norm data in Table~\ref{tab:ConvergenceStudyParabolicBowl}.

\begin{table}[!ht]
\renewcommand{\arraystretch}{1.4}
\centering
\caption{Convergence study for the oscillatory flow in a parabolic basin.}
\begin{tabular}{ccccccc}
\hline
$N$ & $||h-h_{\rm a}||_\infty/||h_{\rm a}||_\infty$ &$\frac{\|e_c\|}{\|e_f\|}$&$||M-M_0||_\infty/||M_0||_\infty$ &$\frac{\|e_c\|}{\|e_f\|}$& $||hu-(hu)_{\rm a}||_\infty$&$\frac{\|e_c\|}{\|e_f\|}$\\
\hline
200 & $1.1\cdot 10^{-3}$&~ & $6.1\cdot10^{-5}$&~ & $5.9\cdot 10^{-2}$&~\\
400 & $3.3\cdot 10^{-4}$&3.33 & $1.7\cdot10^{-5}$ &3.59& $3.4\cdot 10^{-2}$&1.74\\
800 & $7.2\cdot 10^{-5}$ &4.58& $6.9\cdot10^{-6}$ &2.46& $5.2\cdot 10^{-3}$&6.54\\
1600 & $3.9\cdot 10^{-5}$ &1.85& $2.6\cdot10^{-6}$ &2.65& $4.9\cdot 10^{-3}$&1.06\\
\hline
\end{tabular}
\label{tab:ConvergenceStudyParabolicBowl}
\end{table}

\begin{table}[!ht]
\renewcommand{\arraystretch}{1.4}
\centering
\caption{Convergence study for the oscillatory flow in a parabolic basin.}
\begin{tabular}{ccccccc}
\hline
$N$ & $||h-h_{\rm a}||_2/||h_{\rm a}||_2$&$\frac{\|e_c\|}{\|e_f\|}$& $||M-M_0||_2/||M_0||_2$&$\frac{\|e_c\|}{\|e_f\|}$ & $||hu-(hu)_{\rm a}||_2$&$\frac{\|e_c\|}{\|e_f\|}$\\
\hline
200 & $6.1\cdot 10^{-4}$& ~ & $6.1\cdot10^{-5}$&~ & $3.9\cdot 10^{-1}$&~\\
400 & $1.5\cdot 10^{-4}$&4.07 & $1.6\cdot10^{-5}$&3.81 & $1.4\cdot 10^{-1}$&2.79\\
800 & $3.8\cdot 10^{-5}$&3.95 & $6.8\cdot10^{-6}$&2.35 & $5.1\cdot 10^{-2}$&2.75\\
1600 & $1.1\cdot 10^{-5}$&3.45 & $2.6\cdot10^{-6}$&2.61 & $2.1\cdot 10^{-2}$&2.43\\
\hline
\end{tabular}
\label{tab:ConvergenceStudyParabolicBowl2}
\end{table}

\subsubsection{Tsunami run-up on a sloping beach}

The run-up of waves on a sloping beach is a classical test case for numerical schemes for the shallow-water equations within the area of tsunami modeling. As in the case of the parabolic bowl, there exists an analytical solution for this test case, which was first found in the seminal paper~\cite{carr58a}.

\begin{figure}[!ht]
\centering
\begin{subfigure}[b]{0.45\textwidth}
  \centering
  \includegraphics[width=\linewidth]{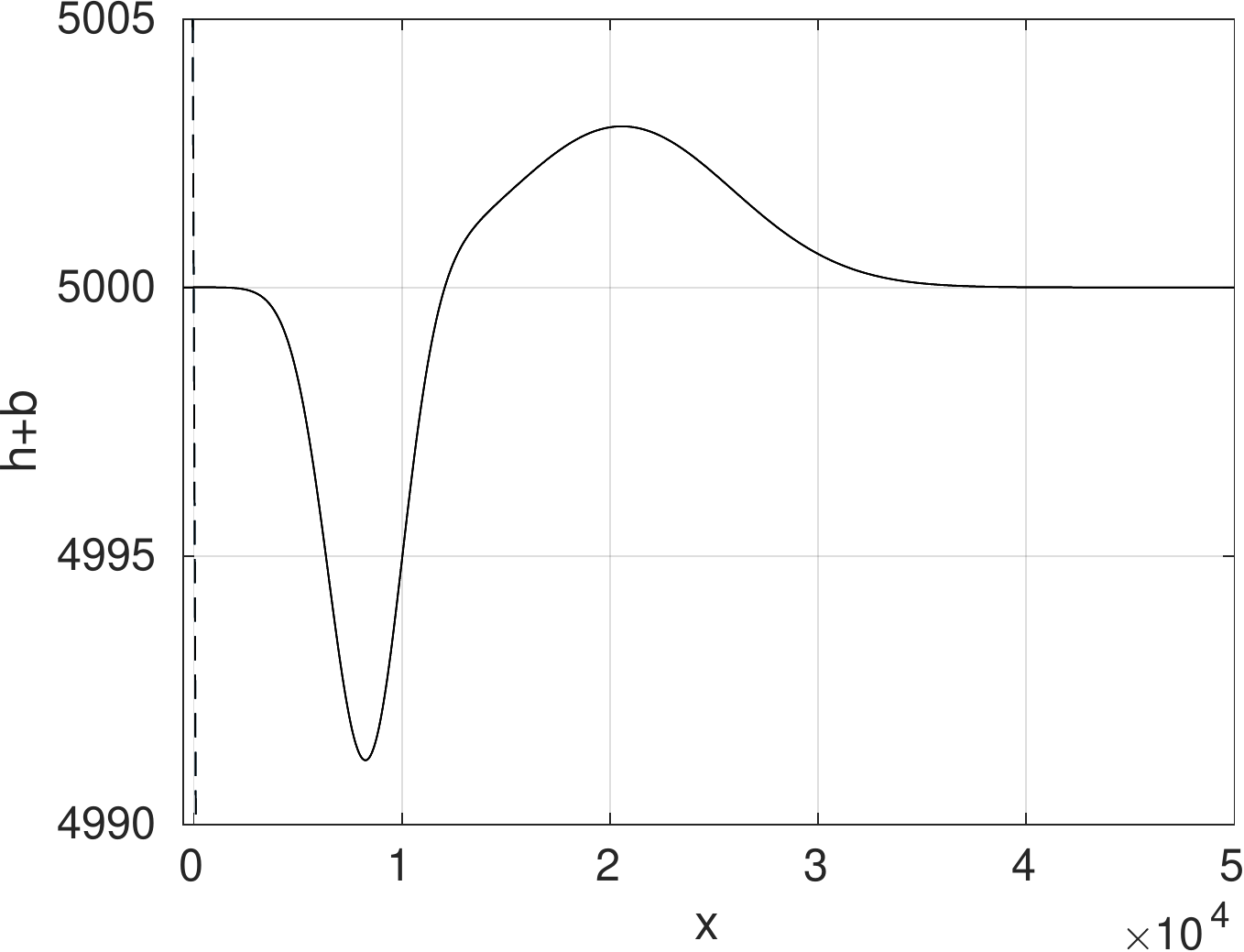}
\end{subfigure}\qquad
\begin{subfigure}[b]{0.45\textwidth}
  \centering
  \includegraphics[width=\linewidth]{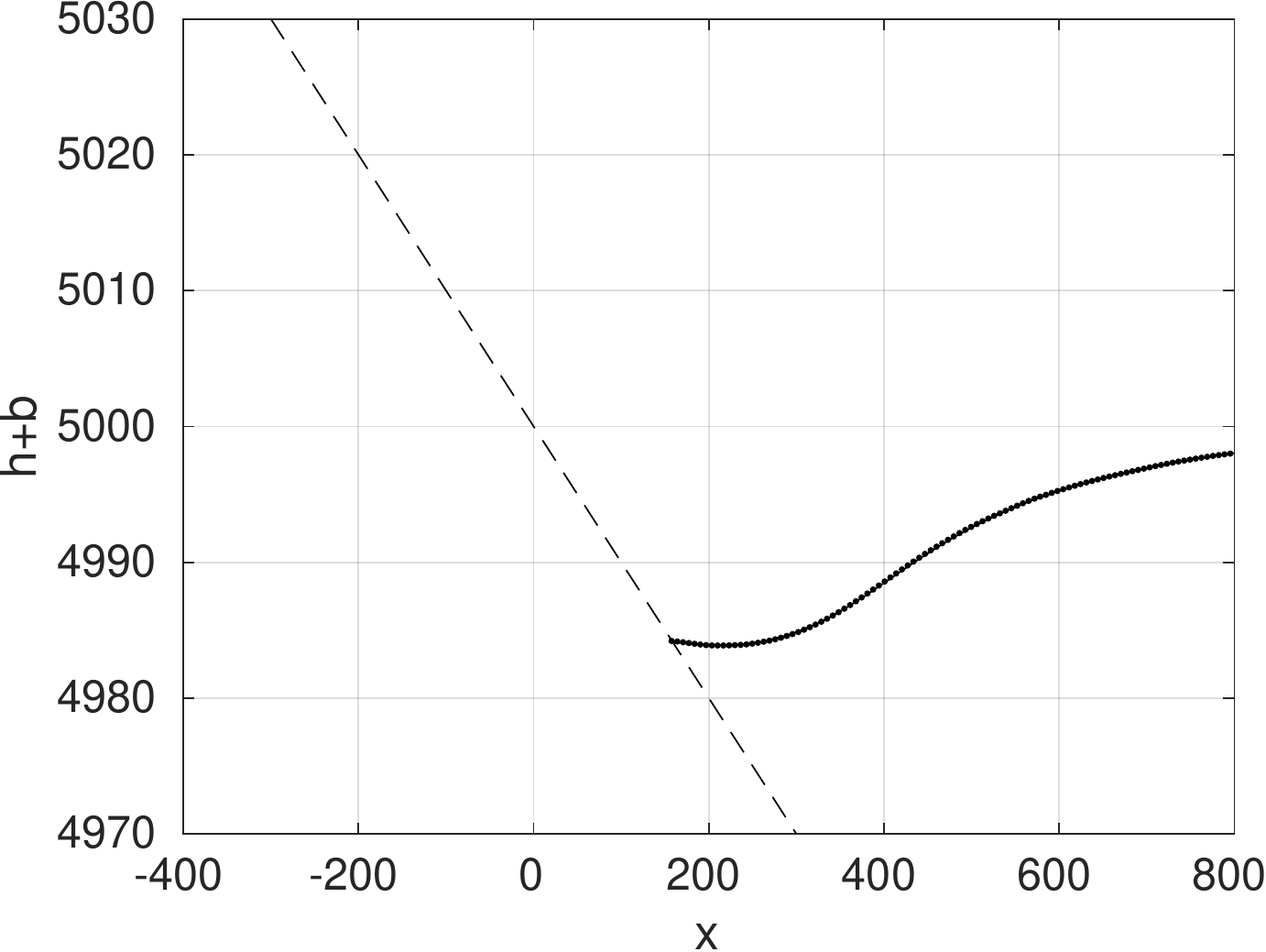}
\end{subfigure}\\
\begin{subfigure}[b]{0.45\textwidth}
  \centering
  \includegraphics[width=\linewidth]{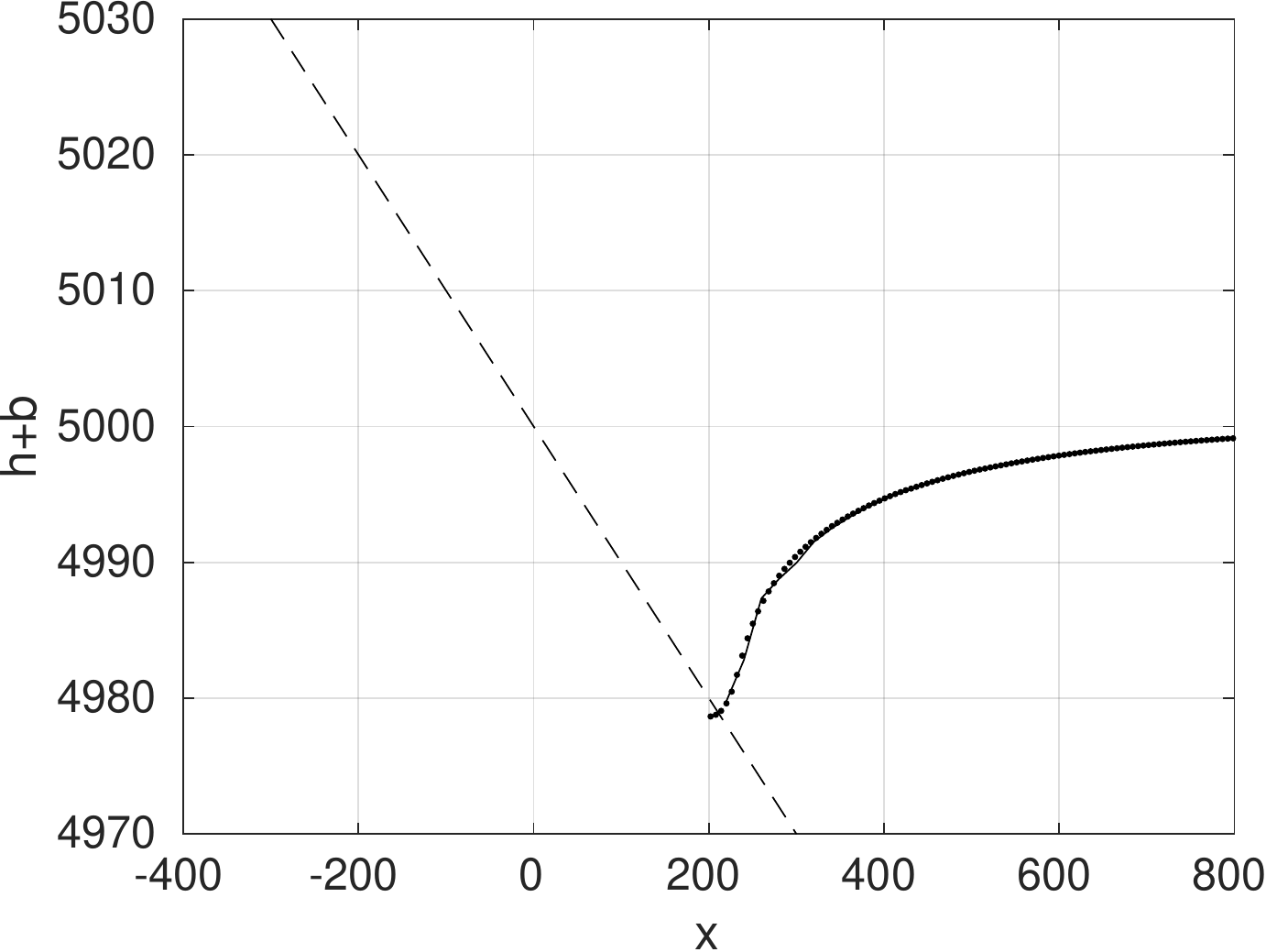}
\end{subfigure}\qquad
\begin{subfigure}[b]{0.45\textwidth}
  \centering
  \includegraphics[width=\linewidth]{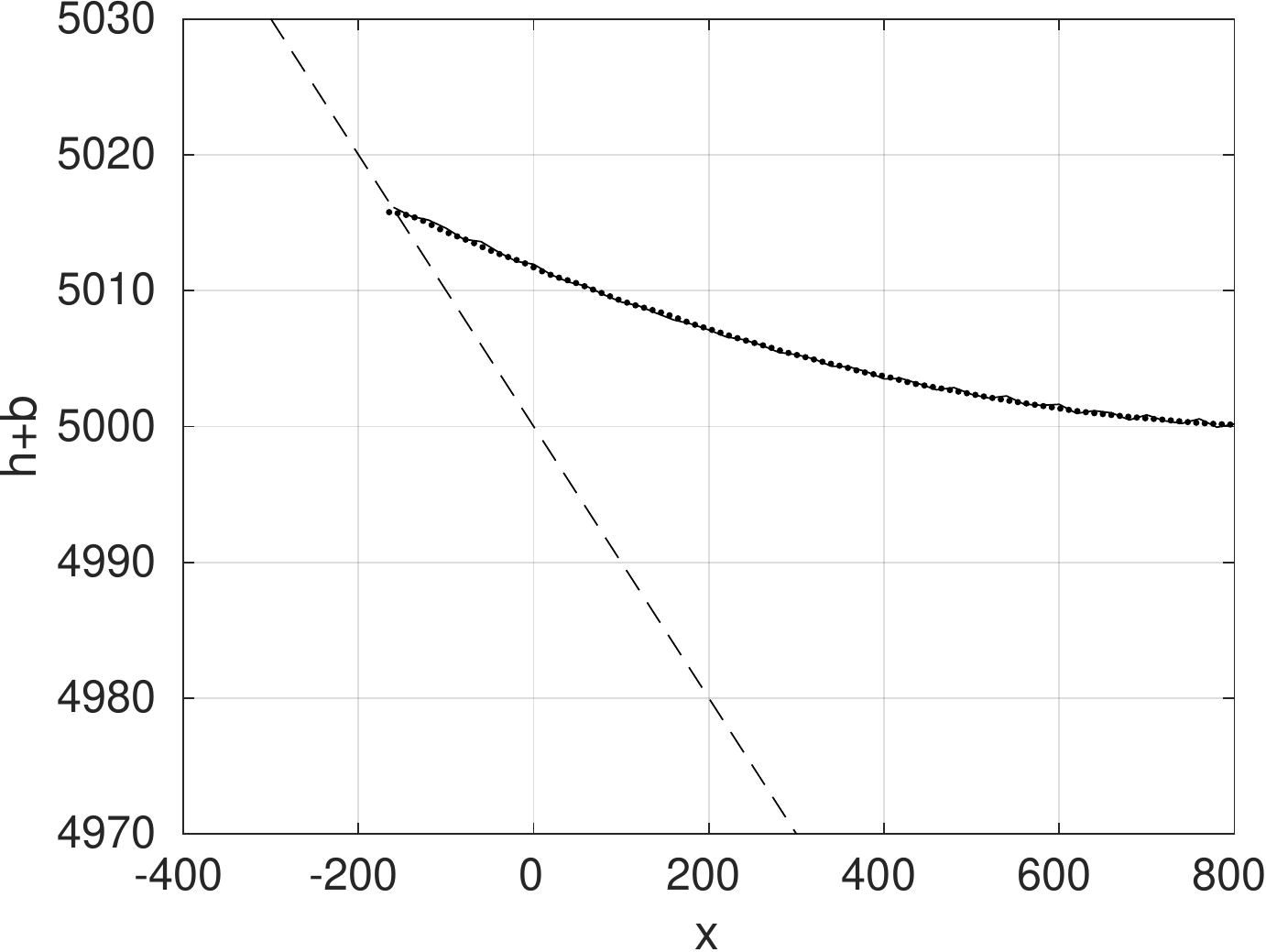}
\end{subfigure}
\caption{\footnotesize{Snapshots of the numerical solution (solid
    line) and analytical solution (dotted line) of the surface
    elevation for the sloping beach bathymetry (dashed line).  Top
    left shows the entire computational domain at time $t=0$, while
    the inundation area is shown for $t=160$ (top right), $t=175$ (lower left) and $t=220$ (lower right).}}
\label{fig:SlopingBeachSolution}
\end{figure}

Here, we follow~\cite{vate15a} and consider the computational domain
$\Omega=[-500,50\ 000]$, with the bottom topography being defined as
$b=5000-0.1x$. Since we will consider the solution at times $t=160$, $t=175$ and $t=220$, simple reflecting boundary conditions can be employed at the right boundary since the reflected waves cannot travel to the sloping beach in that time. We choose $\Delta x=20$ with a time step of $\Delta t=0.025$. The extrapolation parameter was set to $
\delta=0.1$. The initial condition, numerical solutions and exact solutions at the sampling times are displayed in Figure~\ref{fig:SlopingBeachSolution}, which demonstrates that the inundation process is correctly approximated by the numerical model.

\subsection{Two-dimensional benchmarks}

We consider three two-dimensional benchmarks, again the lake at rest,
but on a non-uniform mesh, flow around a conical island, and the
Monai Valley experiment. Again, the well-balanced RBF-FD
discretization described in Section~\ref{sec:RBFFDDiscretization} is
used. The RBF-FD discretization again uses the Gaussian RBF with shape
parameter $\epsilon=0.1/\sqrt{\Delta x^2+\Delta y^2}$, where $\Delta
x$ and $\Delta y$ are the spacings in $x$- and $y$-direction. A total
of nine nearest neighbors are used for the derivative approximation
(which again includes each node as the center of the stencil itself),
unless otherwise indicated. A two-dimensional normalized Gaussian
filter over these nine nearest neighbors is used to construct the
averaging matrix, $\mathrm M$.

\subsubsection{Two-dimensional lake at rest}

\begin{figure}[!ht]
	\centering
	\begin{subfigure}[b]{0.45\textwidth}
		\centering
		\includegraphics[width=\linewidth]{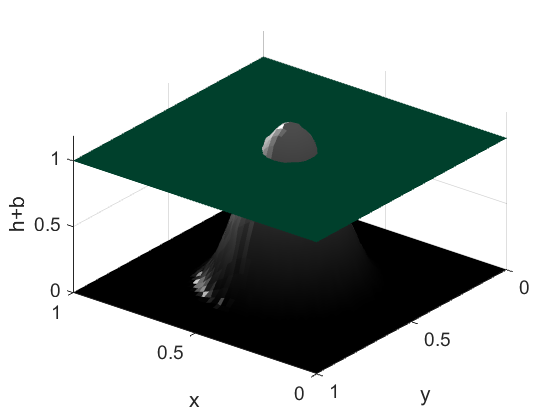}
	\end{subfigure}\qquad
	\begin{subfigure}[b]{0.45\textwidth}
		\centering
		\includegraphics[width=\linewidth]{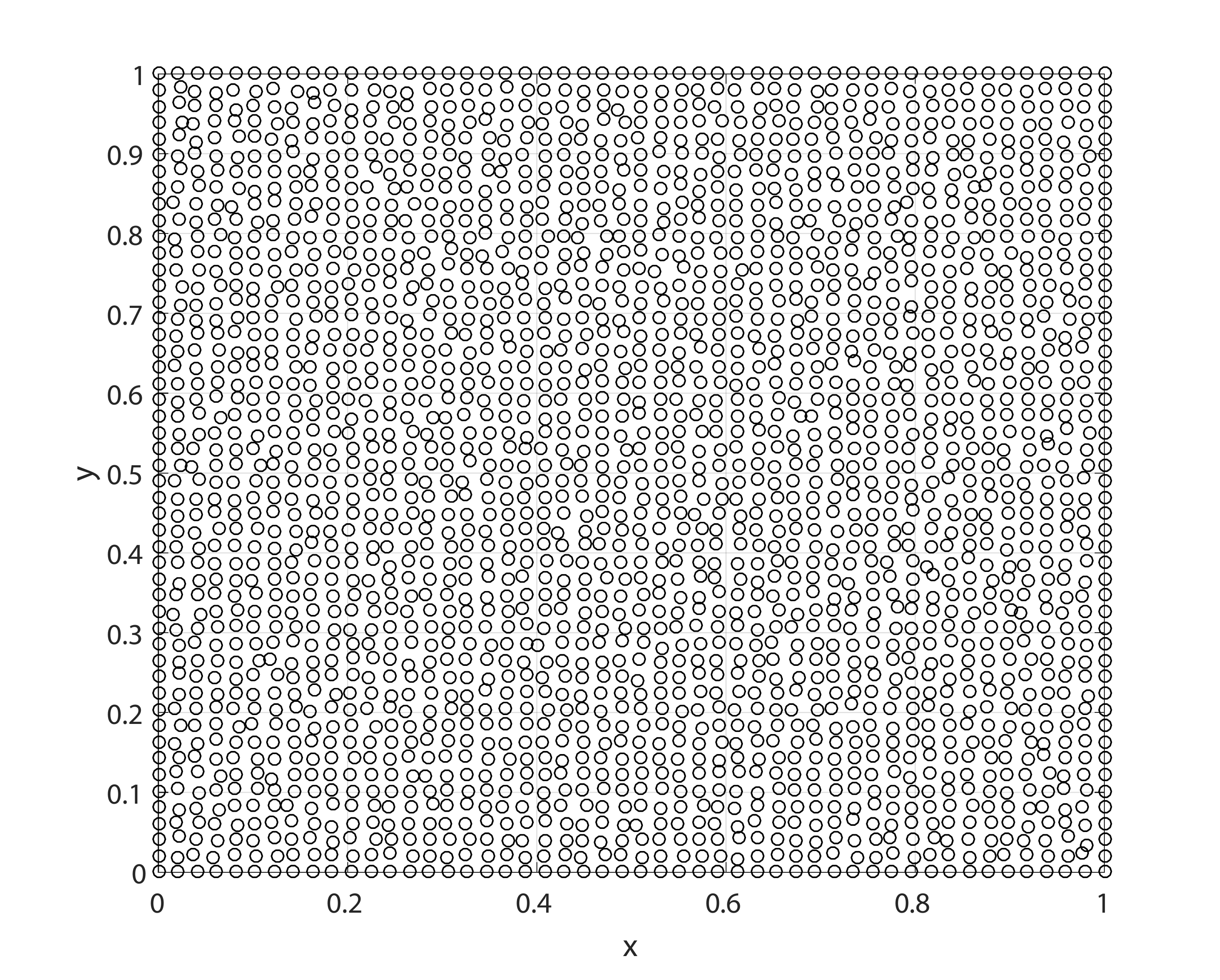}
	\end{subfigure}
	\caption{\footnotesize{Initial surface elevation for the lake at rest solution with smooth bottom topography (left); node distribution for the two-dimensional \textit{lake at rest} test case.}}
	\label{fig:LakeAtRest2d}
\end{figure}

\begin{figure}[!ht]
	\centering
	\begin{subfigure}[b]{0.45\textwidth}
		\centering
		\includegraphics[width=\linewidth]{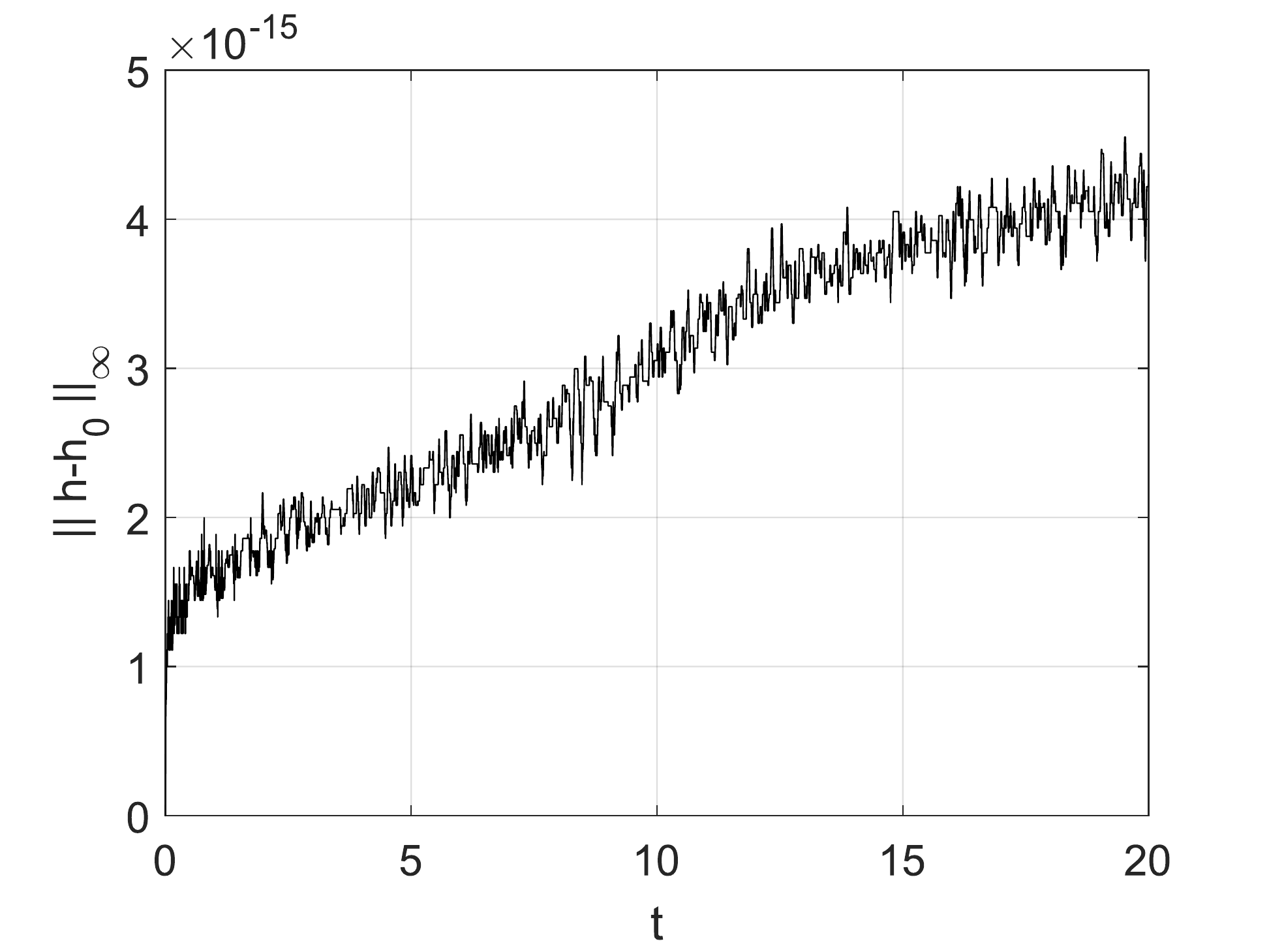}
	\end{subfigure}\qquad
	\begin{subfigure}[b]{0.45\textwidth}
		\centering
		\includegraphics[width=\linewidth]{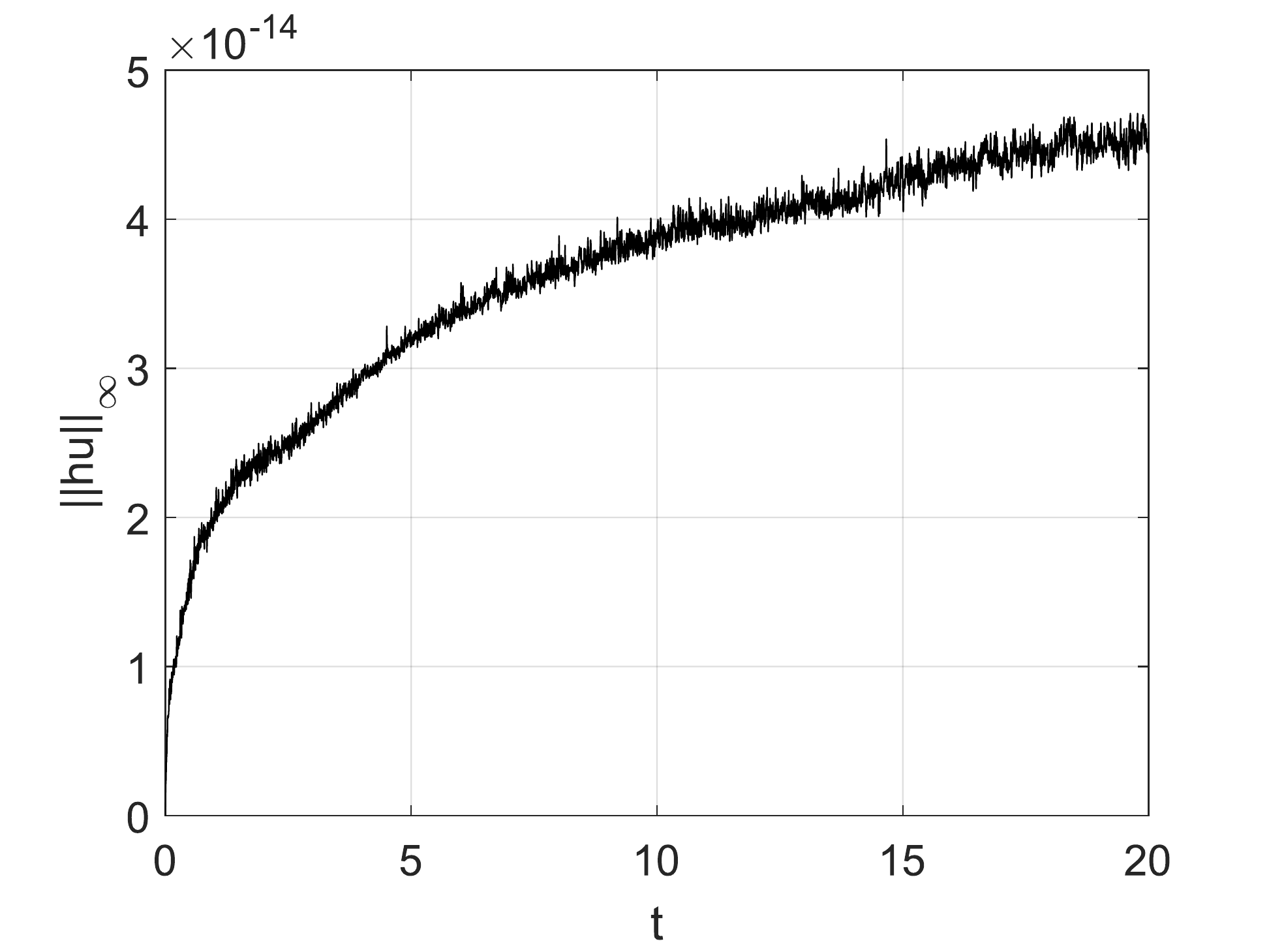}
	\end{subfigure}
	\caption{\footnotesize{$l_\infty$-error time series for the surface elevation (left) and momentum (right) for the lake at rest solution.}}
	\label{fig:LakeAtRest2dErrors}
\end{figure}

We repeat the smooth lake at rest test case in two dimensions but, in
addition, we use a non-uniform mesh. We consider the domain $\Omega =
[0,1]^2$ covered by $n=2500$ nodes. To demonstrate that our scheme is
meshless, we start from a uniform $50\times 50$ mesh of the unit
square and add $(0.1\Delta x \mathcal N(0,1), 0.1\Delta y \mathcal N
(0,1))$ as a disturbance to the coordinates of each grid point, where
$\Delta x = \Delta y = 0.02$ and $\mathcal N(0,1)$ is a normally distributed random variable with zero mean and variance one. We use reflecting boundary conditions.
The bottom topography is given by
$$
b(x,y)=b_s(\|(x-0.5,y-0.5)\|_2)
$$
where $b_s$ is defined in Equation (\ref{bLakeatrestSmooth}).
Again we use $h=\max(0,1-b(x, y))$ and $u=0$ as initial conditions. The time step is $\Delta t=0.0015$ and the final integration time is $t=20$. The extrapolation parameter is set to $\delta=0.05$.
In this case, we use 25 neighbours for the derivative approximation,
to ensure sufficiently accurate approximations on the non-uniform mesh.
\\~\\
The initial surface elevation and node distribution can be seen in
Figure \ref{fig:LakeAtRest2d}. Figure \ref{fig:LakeAtRest2dErrors}
shows the errors in the computed solution over time. 

\subsubsection{Flow around a conical island}

\begin{figure}[!htbp]
	\centering
	\includegraphics[width=0.4\textwidth]{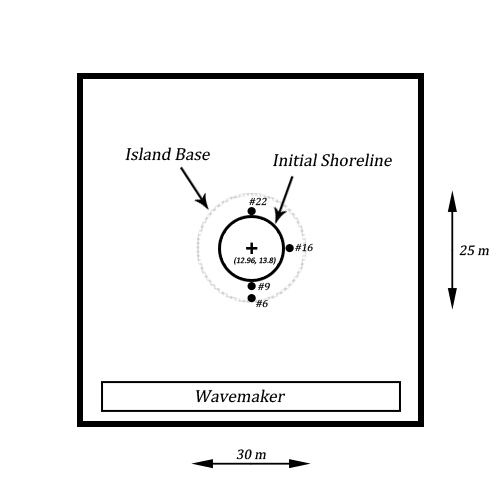}
	~~~ ~~~
	\includegraphics[width=0.4\textwidth]{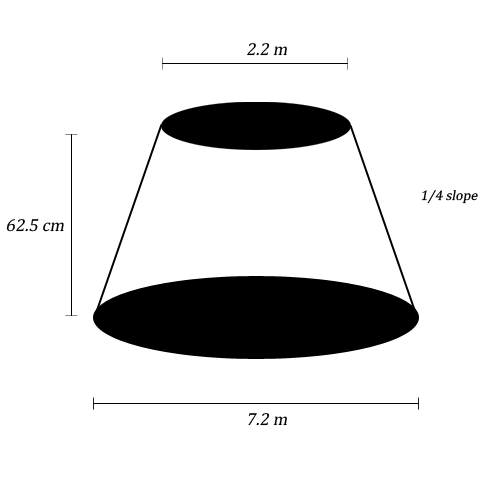}
	\caption{Setup of the experiment and properties of the conical island. Positions of the gauges: (9.36, 13.8)\#6; (10.72, 13.8)$\#9$; (12.96, 11.53)$\#16$; (15.2, 13.8)$\#22$  }\label{conicalIsandSetup}
\end{figure}

\begin{figure}[!htbp] 
	\centering
	\begin{subfigure}[b]{\textwidth}
		\centering
		\includegraphics[width=\linewidth]{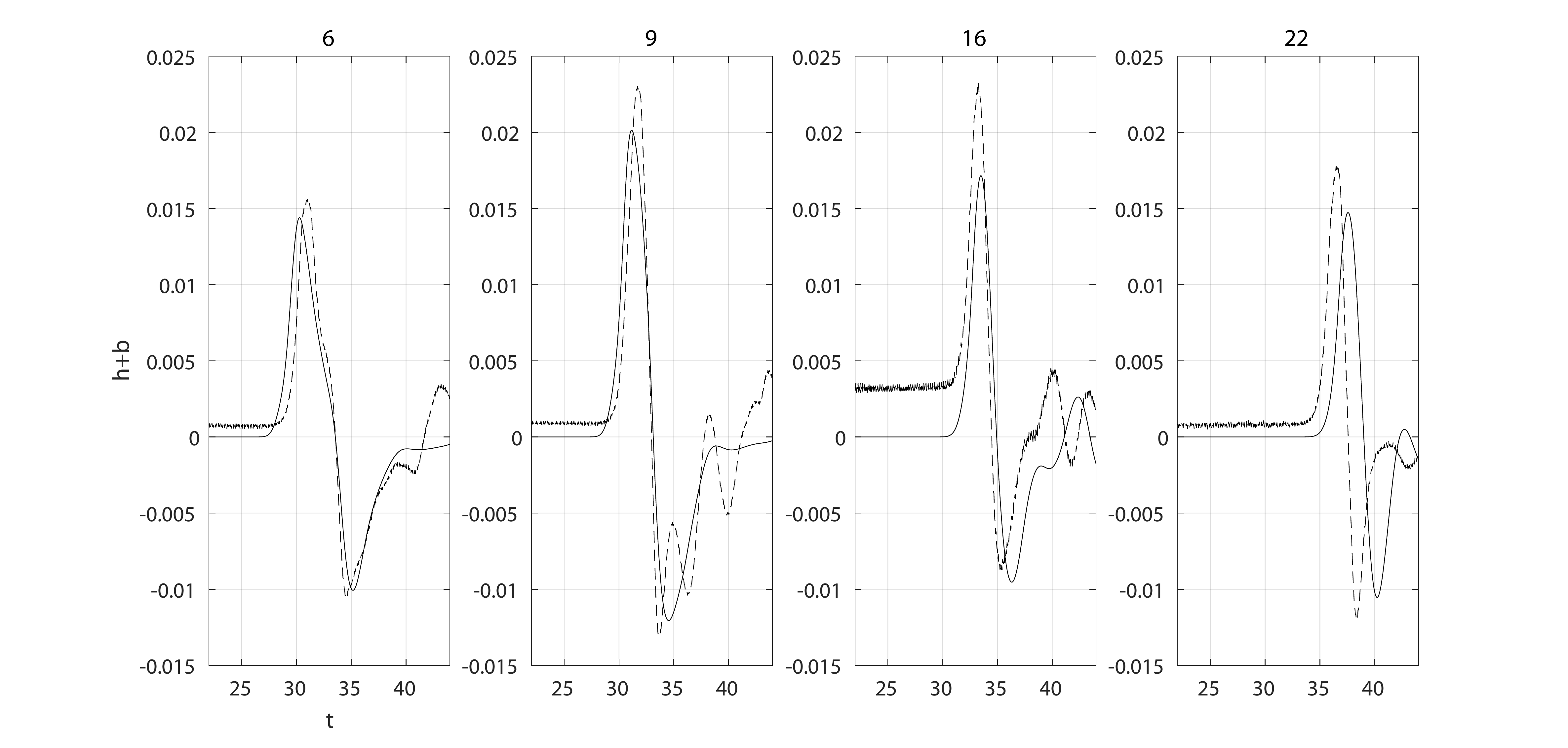}
	\end{subfigure}\qquad
	\begin{subfigure}[b]{\textwidth}
		\centering
		\includegraphics[width=\linewidth]{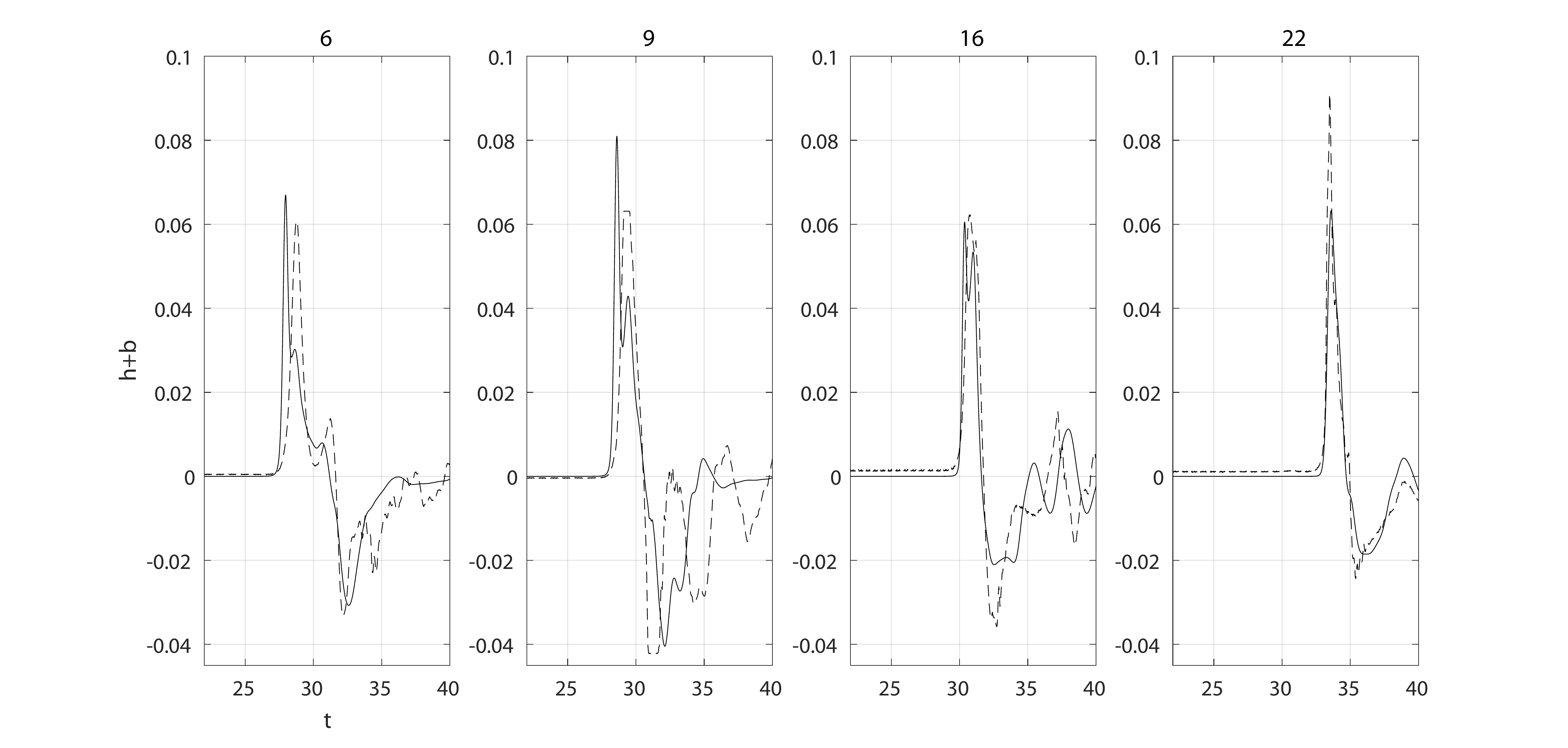}
	\end{subfigure}
	\caption{\footnotesize{ Comparison of measured (dashed) and
            simulated (solid) water elevation at the gauges for case A
            (above) and case C (below) for the conical island.}}
	\label{fig:conicalislandGages}
\end{figure}

\begin{figure}[!htbp] 
	\centering
	\begin{subfigure}[b]{0.45\textwidth}		
		\includegraphics[width=\linewidth]{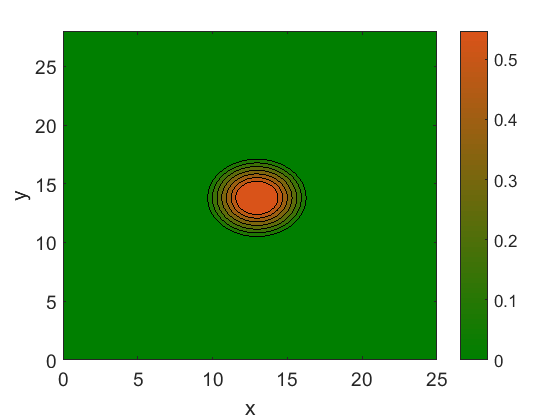}
	\end{subfigure}\\
	\begin{subfigure}[b]{0.45\textwidth}
		\centering
		\includegraphics[width=\linewidth]{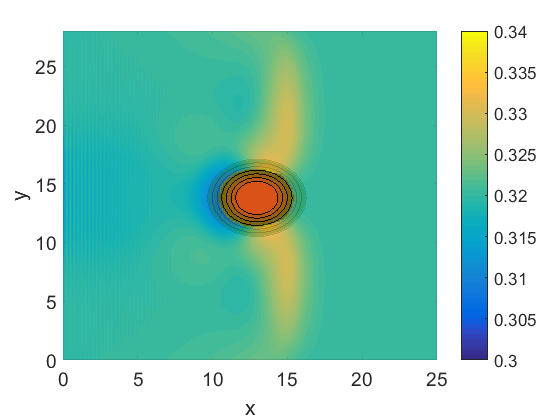}
	\end{subfigure}\qquad
	\begin{subfigure}[b]{0.45\textwidth}
		\centering
		\includegraphics[width=\linewidth]{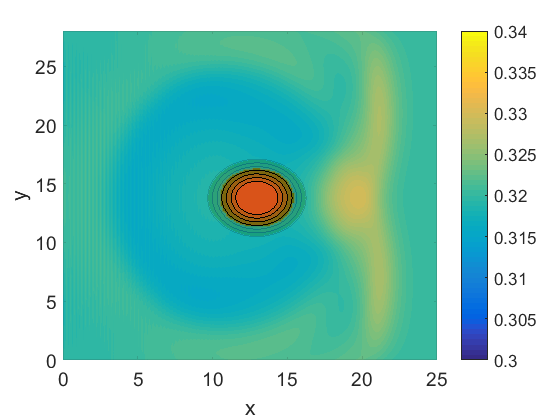}
	\end{subfigure}	
	\begin{subfigure}[b]{0.45\textwidth}
		\centering
		\includegraphics[width=\linewidth]{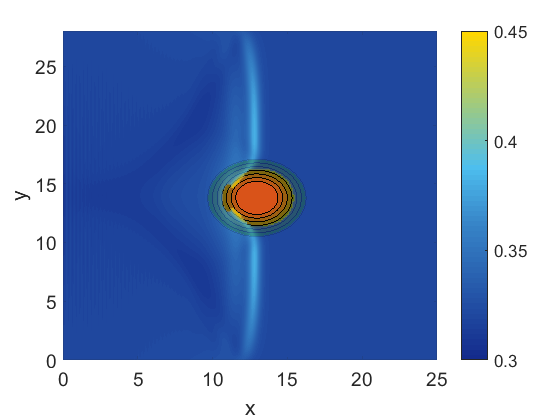}
	\end{subfigure}\qquad
	\begin{subfigure}[b]{0.45\textwidth}
		\centering
		\includegraphics[width=\linewidth]{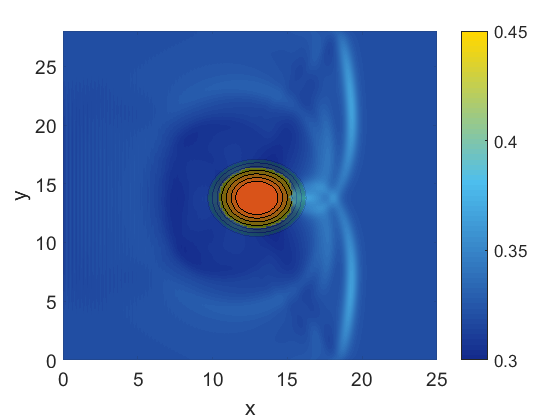}
	\end{subfigure}
	\caption{\footnotesize{Image at the top shows the initial water elevation and the height of the island. Snapshots of the computed solution
            ($h+b$) for the conical island test case at times $t=12$
            and $t=16$ for case A (above) and at times $t=10$ and
            $t=14$ for case C (below) for the conical island.}}
	\label{fig:conicalislandSnap}
\end{figure}

\begin{figure}[!htbp] 
	\centering
\includegraphics[width=\linewidth]{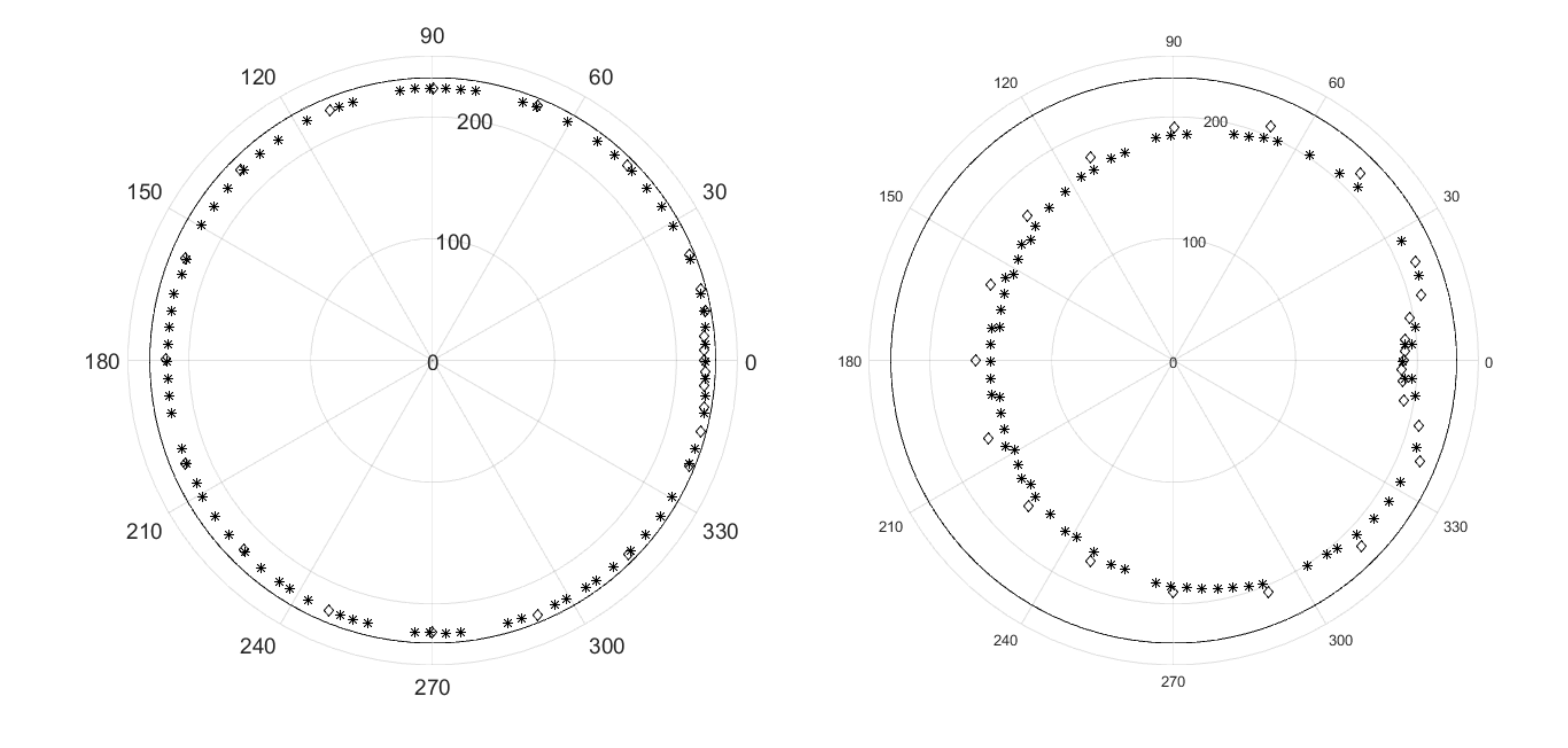}
	\caption{\footnotesize{Run-up for case A (left) and case C
            (right), measured data ($\Diamond$) and computed solution
            ($*$) for the conical island example.}}
	\label{fig:conicalislandRunUp}
\end{figure}

The flow around a conical island is another classical benchmark test for tsunami models, following the experimental study described in~\cite{brig95a}; see also~\cite{liu95a}. 
 The experiment is an idealization of the 1992 Flores Island tsunami run-up on Babi Island. 
The setup of the experiment is a 25m long and 30m wide basin with a flat ground. The border of the basin absorbs the incoming wave, therefore there are no reflections. 
\\
The origin of this conical island is located at $x=12.96$m and $y=13.8$m.
In Figure \ref{conicalIsandSetup}, the setup, shape and properties of the island can be seen. The initial water depth is $h_0=0.32$m. The $y$-axis is parallel to the wavemaker, which generates solitary waves. 
We will compare the data of 4 of the 27 gauges originally used in the experiment, to measure the surface wave elevations. The position of the four gauges are indicated in Figure \ref{conicalIsandSetup}.\\
We choose $\Delta x=0.125, \Delta y=0.14$ and $\Delta t=0.02$. Then we
simulate case A, with  height-to-depth ratio $H=0.04$,
and case C, with $H=0.18$. The recorded water heights at gauges 6, 9, 16 and 22 can be seen in Figure 
\ref{fig:conicalislandGages}. Figure \ref{fig:conicalislandSnap} shows snapshots of the propagating wave for both cases. The maximal horizontal run-up agrees well with the measured data of the experiment, which can be seen in Figure \ref{fig:conicalislandRunUp}.

\pagebreak
\subsubsection{Monai Valley experiment}

The Monai Valley experiment~\cite{mats01a} is a model of the 1993 Okushiri tsunami,  which caused an extreme 32 meter run-up in the Monai Valley on Okushiri island. The purpose of this experiment is to recreate the run-up in a 1/400 model of the relevant part of Okushiri island, see also~\cite[Chapter 11]{liu08a} for further details.

\begin{figure}[!htbp] 
\centering
\begin{subfigure}[b]{0.45\textwidth}
	\centering
 \includegraphics[width=\linewidth]{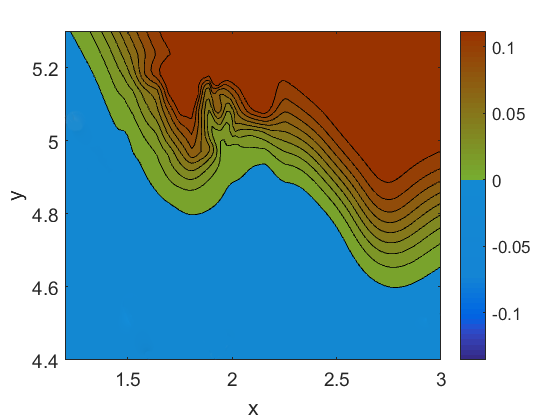}
\end{subfigure}\\
\begin{subfigure}[b]{0.45\textwidth}
  \centering
 \includegraphics[width=\linewidth]{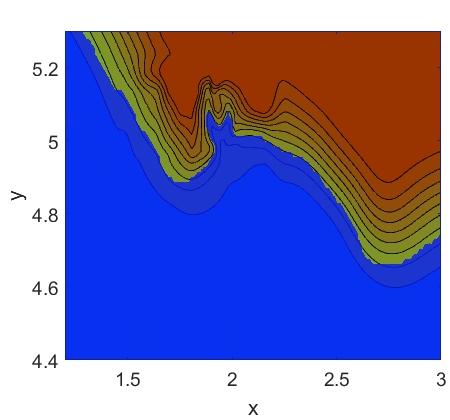}
\end{subfigure}\qquad
\begin{subfigure}[b]{0.45\textwidth}
  \centering
  \includegraphics[width=\linewidth]{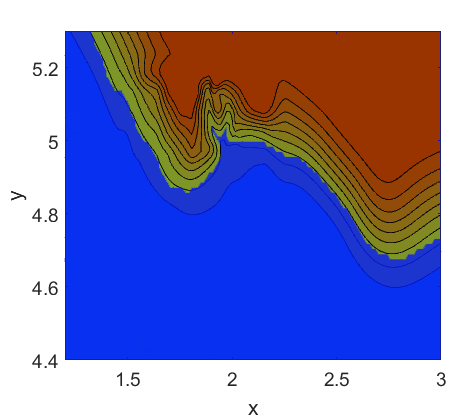}
\end{subfigure}\\
\begin{subfigure}[b]{0.45\textwidth}
  \centering
 \includegraphics[width=\linewidth]{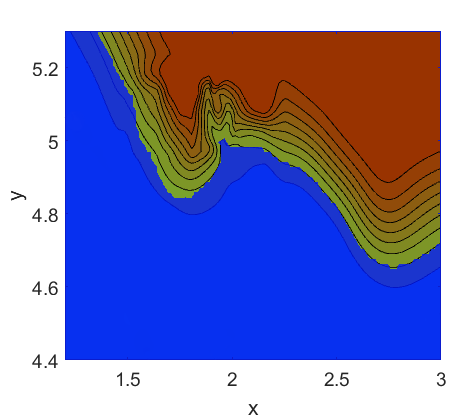}
\end{subfigure}\qquad
\begin{subfigure}[b]{0.45\textwidth}
  \centering
\includegraphics[width=\linewidth]{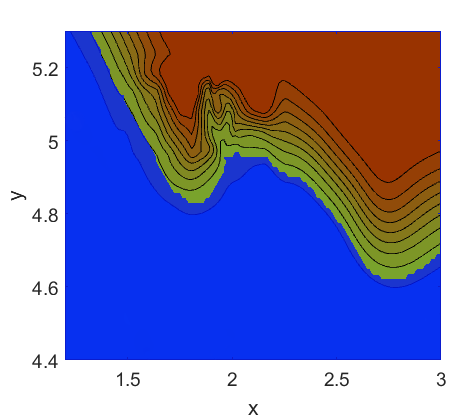}
\end{subfigure}
\caption{\footnotesize{Image at the top shows the initial water elevation and the height of the valley. Snapshots of the computed solution ($h+b$) for the Monai Valley test case at times $t=16.2$ (top left) (time of maximal run-up), $t=17$ (top right), $t=18$ (lower left) and $t=19$ (lower right).}}
\label{fig:MonaiValleyWaterHeight}
\end{figure}

\begin{figure}[!htbp] 
  \centering
 \includegraphics[width=\linewidth]{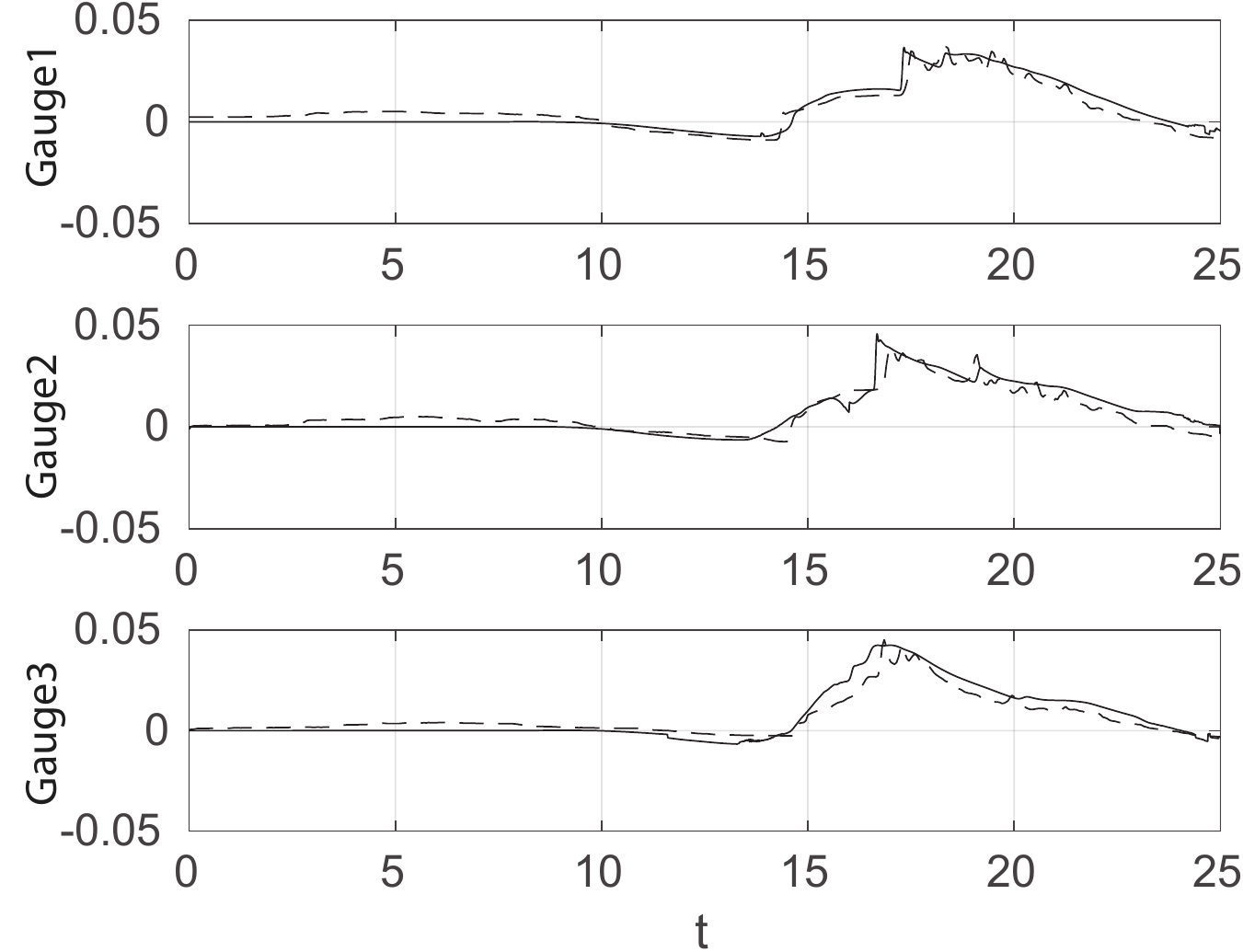}
\caption{\footnotesize{Gauges 1,2,3 for the Monai Valley test case.}}
\label{fig:MonaiValleyGages}
\end{figure}

We discretize the domain $\Omega=[0,5.488]\times[0,3.402]$ with a regular mesh with step sizes $\Delta x=\Delta y=0.014$. Reflective boundary conditions are employed everywhere except at $x=0$, where the incident wave is prescribed up to $t=22.5$. For $t>22.5$ we use open boundary conditions at $x=0$. Water levels are recorded at the three points $(4.521,1.196)$, $(4.521,1.696)$ and $(4.521,2.196)$, which correspond to gauges 1, 2 and 3 of the experimental setup at which points measurements of the water height are provided. We integrate the shallow-water equations until $t=25$, which is long enough to record the maximum run-up which occurs at approximately $t=20$ at the reference locations. The time step in the simulation was set to $\Delta t=0.01$. All experimental data as well as the incident wave profile were obtained from~\cite{mona17a}.
\\
The evolution is illustrated by four snapshots in Figure \ref{fig:MonaiValleyWaterHeight}.
The recorded water heights at the gauges is in good accordance with the experimentally recorded values, see Figure \ref{fig:MonaiValleyGages}.

\section{Conclusions}\label{sec:ConclusionsTsunami}

In this paper, we have proposed a novel numerical procedure for
solving the shallow-water equations, which is suitable for tsunami modeling. In particular, both the well-balanced numerical scheme and the inundation algorithm are based on radial basis functions. This makes the model truly meshless and, hence, capable of employing variable resolution as well as operating on arbitrary coastal regions, without the need to use an underlying orthogonal mesh. First benchmark tests demonstrate the competitiveness of the proposed methodology both in the one- and two-dimensional setting.

A main defining characteristic of tsunamis is that they occur on a multitude of spatial and temporal scales, which can pose considerable challenges to numerical solvers for the shallow-water equations. An advantage of the RBF methodology is that is can be easily adopted to both arbitrary geometries (e.g.\ the plane and the sphere) and arbitrary nodal layouts, both features important for the far- and near-field modeling of tsunamis. While, in the present work, we have concentrated on the near coast propagation of tsunamis as well as coastal inundation, in future work we will combine the methodology proposed here with a global scale tsunami propagation model. Suitable candidates for numerical models have already been proposed in~\cite{flye09Ay,flye12a}, which we will adopt to be able to handle arbitrary sea bottom topographies.

With the exception of the two-dimensional lake at rest solutions, all
other benchmark tests were carried out on a regular, orthogonal
mesh. This was done in order to facilitate comparison with other
numerical models for the same benchmark problems that are usually done
on an orthogonal mesh as well. A full demonstration of the meshless
capabilities of the proposed shallow-water discretization, as well as
more complicated real-world tsunami simulations, is a subject for
future work.

One potential problem of the RBF-FD methodology as used in the present
paper is that the resulting schemes are not exactly mass
conserving. While we have numerically verified that the error in mass
conservation is purely oscillatory (and showing no spurious growth or
decay), in order for the meshless methodology to be competitive with
standard finite volume or discontinuous Galerkin methods (which both
typically preserve mass) it will be essential to develop a meshless
methodology that is conservative when applied to hyperbolic
conservation laws. We will consider this issue in future investigations.

\section*{Acknowledgments}

This research was undertaken, in part, thanks to funding from the
Canada Research Chairs program and the NSERC Discovery Grant
program. 
J.B. acknowledges support by the Cluster of Excellence CliSAP (EXC177), Universit\"at Hamburg, funded by the German Science Foundation (DFG). Further support by Volkswagen Foundation (Project ASCETE, AZ 88 470) is acknowledged.
The authors thank Stefan Vater for valuable advice on setting up and interpreting test cases.

\footnotesize\setlength{\itemsep}{0ex}

\end{document}